\pdfoutput=1
\documentclass[conference]{IEEEtran}
\IEEEoverridecommandlockouts
\usepackage{cite}
\usepackage{amsmath,amssymb,amsfonts}
\usepackage{algorithmic}
\usepackage[linesnumbered, ruled]{algorithm2e}
\SetKwRepeat{Do}{do}{while}%
\usepackage{algorithmic}
\usepackage{graphicx}
\usepackage{booktabs}
\usepackage{textcomp}
\usepackage{url}
\usepackage{hyperref}
\usepackage{xcolor}
\def\BibTeX{{\rm B\kern-.05em{\sc i\kern-.025em b}\kern-.08em
    T\kern-.1667em\lower.7ex\hbox{E}\kern-.125emX}}
\begin{document}

\title{CloudQC: A Network-aware  Framework for Multi-tenant Distributed Quantum Computing
}
% sb cao ni ma 
\author{
\IEEEauthorblockN{Ruilin Zhou\IEEEauthorrefmark{1}, Yuhang Gan\IEEEauthorrefmark{1}, Yi Liu, and Chen Qian}
%\IEEEauthorblockA{\textit{Department of Computer Science and Engineering} \\
\textit{University of California, Santa Cruz}\\
%\textit{Santa Cruz, USA} %\\
%\{ygan11, xzhan330, rzhou39, yliu634, cqian12\}@ucsc.edu}
}

\maketitle
\footnotetext[1]{Authors contributed equally to this research.}
\begin{abstract}
Distributed quantum computing (DQC) that allows a large quantum circuit to be executed simultaneously on multiple quantum processing units (QPUs) becomes a promising approach to increase the scalability of quantum computing. It is natural to envision the near-future DQC platform as a multi-tenant cluster of QPUs, called a Quantum Cloud. However, no existing DQC work has addressed the two key problems of running DQC in a multi-tenant quantum cloud: placing multiple quantum circuits to QPUs and scheduling network resources to complete these jobs. This work is the first attempt to design a circuit placement and resource scheduling framework for a multi-tenant environment. The proposed framework is called CloudQC, which includes two main functional components, circuit placement and network scheduler, with the objectives of optimizing
both quantum network cost and quantum computing time. Experimental results with real quantum circuit workloads show that CloudQC significantly reduces the average job completion time compared to existing DQC placement algorithms for both single-circuit and multi-circuit DQC. We envision this work will motivate more future work on network-aware quantum cloud. %in networking. 
%We proposed QCloud with the following designs: 1)We proposed a quantum cloud architecture that combines the power of quantum networks and quantum computing devices and we use such infrastructure to perform distributed quantum computing jobs. 2) We proposed a scheduling framework which can first map quantum computing jobs to quantum computing devices with a focus on improving resources utilization and per-job performance. 3) After finding placement for each job, we used our proposed flow scheduler to allocate resources for inter-qpu communication to optimize per-job performance. We evaluate our work on various real-world quantum circuits benchmarks, and generated different workloads to evaluate our system. 

% Previous compiling works on quantum computing mainly focus on compile and schedule single circuit. 
% In this paper we envision such a quantum cloud framework, and proposed QCloud, the first scheduling framework in quantum cloud that 1) it first maps concurrent incoming tasks to quantum computing devices in quantum cloud with a focus on resources utilization and performance. 2) then it schedules inter-commnucation between different quantum computing devices to achieve the best performance. 
\end{abstract}

%\begin{IEEEkeywords}
%component, formatting, style, styling, insert
%\end{IEEEkeywords}

\section{Introduction}
% Inrtoduction 主要就是写之前的quantum conputing 大概什么样，最近quantum computing的发展（IBM 1000+， IONQ All connecred quantum computing， QuEra logical qubits)，DQC 和multi-programming提出，和最近quantum的一些发展都让一个大规模 的 quantum cloud变得可能。 那么有很多问题需要解答，如果真的有这样的quantum cloud他应该有哪些部分组成，这是第一个要回答的问题。 
% 另外要提出现在quantum cloud的问题。首先现在大部分quantum供应商提供services都是以cloud的方法， IBM的quantum cloud long waiting time, resources under utilization. 这些加上之前提的那些trend都展示出需要：
% 一个新的quantum cloud的architecture
% 一个新的scheduling strategy同时考虑到utilziation和performance
% 如果大规模跑，
%   那么需要dqc，dqc需要的scheduling和普通quantum cirucit的 mapping， scheduling 是不同的，所以针对于单个circuit需要新的scheduling 方法
%   另外quantum cloud的计算模型和传统不一样，所以需要新的cloud的 job scheduling的方法a

Quantum computing technologies have shown great potential to solve complex problems and provide speedups such as quadratic speed up in database search \cite{grover1996fast}, and simulations for physical sciences \cite{feynman2018simulating}. %On the other hand, quantum networks can support various applications such as quantum sensor networks \cite{eldredge2018optimal}, synchronization  \cite{vinokur2008superinsulator} and provide more secure communication techniques than classical cryptography\cite{shor2000simple}. 
Current quantum computing techniques are in the Noisy Intermediate-Scale Quantum (NISQ) era \cite{ding2020quantum}, characterized by limited qubits, limited connectivity, and prevalent noise. These challenges greatly hinder the usage of quantum computing since it is commonly assumed that at least millions of qubits are required according to the current qubits error rates to execute a practical quantum algorithm \cite{monroe2022building}. Recent advances show that increasing the number of qubits on a single quantum processor is challenging due to hardware limitations such as crosstalk errors \cite{Bruzewicz2019}, qubit addressability \cite{Bruzewicz2019}, and fabrication difficulty \cite{brink2018challenges}. More importantly, the significance of these challenges usually increases with the size of quantum hardware \cite{wu2022autocomm}. 
%Recent advances in quantum computing and quantum networks, however, are pushing these boundaries in various aspects:1) Better modularity architecture with an increasing number of physical qubits: The most advanced superconducting qubits have increased to more than 1000 qubits utilizing cross-resonance gate technology and the envisioned future modular quantum processor design~\cite{ibm2023roadmap}.2) Advances in quantum error correction: An execution of large circuits with 48 logical qubit systems with up to 280 physical qubits have been proposed~\cite{bluvstein2024logical}. 3) Advances in quantum interconnects: Successful multi-node entanglement demonstrations using NV centers\cite{pompili2021realization}, which enables fast inter-node communication. Besides these advances in these fundamental quantum experiments, various efforts have also been proposed to better utilize quantum technology: 
Instead of waiting for a powerful single quantum computer, a promising approach to increase scalability is distributed quantum computing (DQC) \cite{monroe2014large,wu2022autocomm,wu2023qucomm,cuomo_optimized_2023,mao2023qubit,hermans2022qubit,niu2023low,magnard2020microwave,li2024high}. 
DQC makes use of the resources of multiple quantum computers to perform large quantum computing jobs by separating each job 
onto different quantum computers 
and using inter-processor communication \cite{monroe2014large}. 
Quantum computers working on the same DQC task rely on a quantum network to exchange quantum bits (qubits) \cite{monroe2014large,wu2022autocomm,wu2023qucomm,cuomo_optimized_2023,pompili2021realization,mao2023qubit,niu2023low,magnard2020microwave,li2024high}.  
%Progress has been made from both compiler works\cite{wu2022autocomm,wu2023qucomm,cuomo_optimized_2023} and experiments\cite{pompili2021realization,hermans2022qubit,niu2023low,magnard2020microwave,li2024high}. 

It is natural to envision the near-future QDC platform as a shared cluster of quantum computers \cite{ravi2021quantumcloud,liu2021qucloud}, called a \textit{Quantum Cloud}, similar to today's cloud computing. This is because quantum hardware is more expensive for individual users. For example, industry companies like IBM \cite{ibmquantumcloud}, Microsoft \cite{msquantumcloud}, and NVIDIA \cite{nvquantumcloud} have already launched their quantum cloud services. 
Recently developed quantum multi-programming~\cite{das2019case,liu2021qucloud,niu2023enabling} enables multiple quantum computing jobs (also called as quantum circuits) to be executed on the same quantum computer simultaneously. This technology further allows a multi-tenant quantum cloud that supports DQC to become a reality in the near future. 

One fundamental research problem of DQC is to divide a quantum computing job, represented by a \textit{quantum circuit}, into sub-circuits and place them onto different quantum computers such that each quantum computer has sufficient resources to assign to these sub-circuits  \cite{wu2022autocomm,wu2023qucomm,mao2023qubit,andres-martinez_distributing_2023}. Finding a placement that optimizes inter-quantum-computer communication is crucial because sub-circuits on different quantum computers will cause quantum network communication that spends Einstein-Podolsky-Rosen (EPR) pairs (also called \textit{entanglements}), whose generation is known to be expensive \cite{wu2022autocomm,wu2023qucomm,mao2023qubit,andres-martinez_distributing_2023}. However, the prior solutions \cite{wu2022autocomm,wu2023qucomm,mao2023qubit,andres-martinez_distributing_2023} focus on placing a single circuit, and none considers a multi-tenant cloud environment, where multiple users request their quantum circuits to be executed in the cloud simultaneously. 

We envision that circuit placement in a multi-tenant cloud will become a new line of research because 1) it aligns with the trend of developing a high-throughput, low-latency, quantum cloud; 2) this problem \textit{significantly differs} from existing research on placing one single circuit in DQC and that on placing virtual machines (VMs) in classic cloud computing. 

Circuit placement in a multi-tenant cloud cannot directly use single-circuit solutions \cite{wu2022autocomm,wu2023qucomm,mao2023qubit,andres-martinez_distributing_2023} because, 1) an optimized placement for one circuit could leave difficult remaining resource to place other circuits. Hence, the process of placing one circuit should include the consideration of whether the remaining resource is difficult for other placements; 2) prior solutions do not consider scheduling network resources (EPR pairs) among different circuits, while network resource contention should be a concern in a multi-tenant cloud.

 On the other hand, existing network optimization methods for VM placement in a classic cloud \cite{meng2010VM,ballani2010predictable,Masdari2016VMOverview} do not fit into a quantum cloud either. There are two main reasons: 1) different separations of the same quantum circuit will cause significantly different communication costs, while the network traffic between different VMs will not change a lot by their placement; 2) one key feature of quantum network operations is its probabilistic nature~\cite{van2014quantum}. EPR generation does not always succeed. Hence the quantum communication in DQC could fail even when sufficient resource is allocated. In a classic cloud, network communication is assumed to be successful if the bandwidth allows. 

To our knowledge, there is no existing work that has addressed the above concerns for a multi-tenant quantum cloud. \textbf{This work is the first attempt to design a circuit placement and resource scheduling framework for a multi-tenant quantum cloud. 
Regardless of how the exact quantum computing technologies will evolve, 
the research problem studied in this work is a necessary step for multi-tenant quantum computing platforms.}

The proposed framework is called CloudQC, which includes two main functional components: circuit placement and network scheduler. The circuit placement function allows the quantum cloud to assign resources to quantum circuits as a bunch (at system initialization) or a sequence (when requests from users come sequentially). The placement of one circuit will optimize the remaining resources for other circuits. The network scheduler function assigns network resources (EPR pairs) among different sub-circuits. For important quantum gates, CloudQC will assign them redundant network resources to avoid potential backlogs that could cause long latency for the entire circuit. CloudQC can also be used for single-circuit placement and it improves existing methods. The implementation of our simulation is available to the public~\cite{cloudqc}.
Our main contributions are summarized as follows:

\begin{itemize}
    \item We are the first to model a multi-tenant quantum cloud architecture that supports multiple QDC jobs and optimizes both quantum network cost and quantum computing performance.
      \item With the goal of improving both utilization, throughput, and performance of the quantum cloud, we propose new placement and scheduling methods that allow quantum circuits to be executed on multiple QPUs. %We designed a novel community-detection algorithm to determine the allocation of each job. 
     We design a network scheduler that considers the quantum network's probabilistic nature and allocates communication qubits to each remote quantum gate. 
    \item Simulation results with real quantum circuit workloads show that CloudQC significantly reduces the job completion time compared to existing DQC placement algorithms \textbf{for both single-circuit and multi-circuit DQC}.  
\end{itemize}
The rest of the paper is organized as follows: we first introduce necessary background knowledge. Then, we will introduce how we model components in our proposed quantum cloud and how we model essential quantum operations in our work. 
%\vspace{-1ex}
\section{Background}
%\vspace{-1.5ex}
%\subsection{Qubits and Entanglement}
% \vspace{-1ex}
\textbf{Qubits and entanglement.}
 A quantum bit (qubit) is the fundamental unit in quantum computing and quantum networks. 
 %Properties such as entanglement and superposition make qubits different from classical bits. One classical bit can have either 1 or 0 values, but 
 A qubit can be in the superposition state of 1 and 0 \cite{nielsen2002quantum}. Such a state can expressed mathematically as: 
$|\Psi\rangle=\alpha|0\rangle+\beta|1\rangle$. 
%, here $|\Psi\rangle$ denotes a quantum state and both $|\alpha|$ and $|\beta|$ are complex number. 
Upon the measurement on a qubit, the quantum state of the qubit will become either state $|0\rangle$ or state $|1\rangle$ with probability $|\alpha|^{2}$ or $|\beta|^{2}$. An important feature of the quantum state is called an entanglement of two qubits. 
%An example of entanglement can be expressed as follows:$\frac{1}{\sqrt{2}}(|00\rangle+|11\rangle)$. 
Two entangled qubits are called an EPR pair~\cite{nielsen2002quantum}, which is a fundamental unit in quantum communication. 
%To perform entanglement swapping or teleportation between two remote devices, one remote EPR pair must be distributed on two devices. Most EPR pair preparation includes two steps: first several low-quality EPR pairs are generated, and then some are sacrificed to generate high-quality EPR pairs.

\begin{figure}[t]
\setlength{\abovecaptionskip}{0.cm}
\centerline{\includegraphics[width=0.49\textwidth]{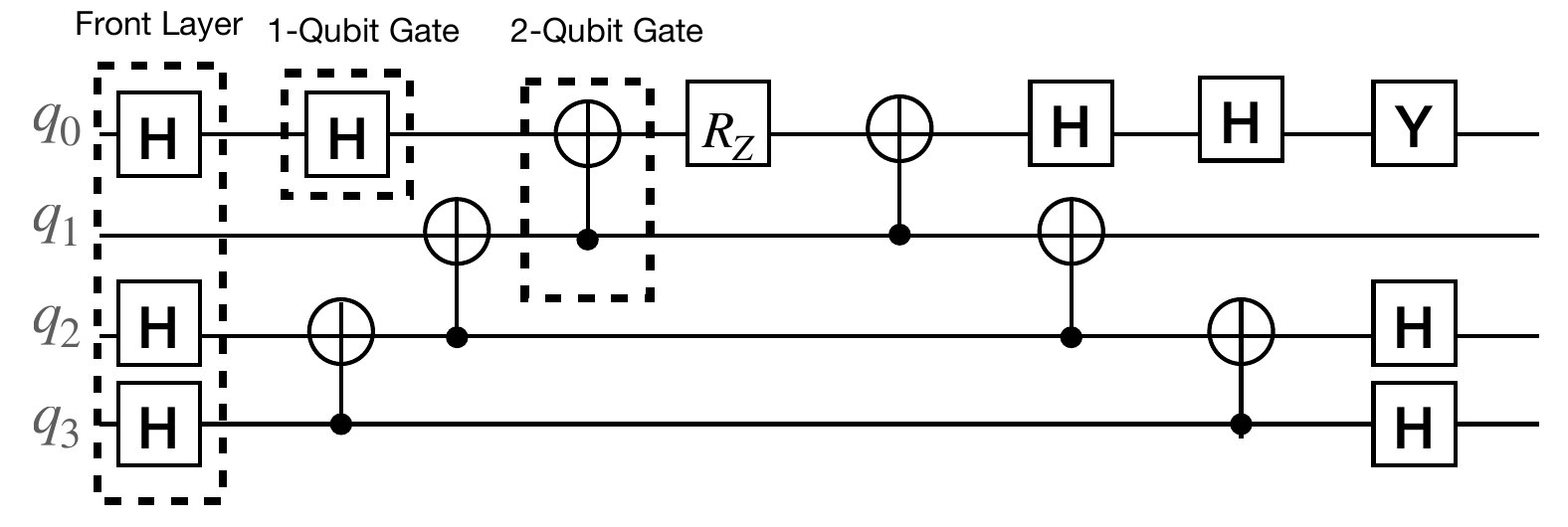}}
\caption{Quantum circuit of a 4-qubit VQE algorithm}

\label{fig:circuit_exmple}
\vspace{-1.5ex}
\end{figure}

%\vspace{-1ex}
%\subsection{Quantum Gates and Quantum Circuits}
%\vspace{-1ex}
\textbf{Quantum gates and quantum circuits.}
Most quantum algorithms adopt a \textit{quantum circuit} model, expressing quantum operations as \textit{quantum gates}. Mathematically, a quantum gate can be considered as a matrix that works on the quantum state. 
%which is a state vector. 
For example, one Hadmard gate (H gate) can be represented by a unitary matrix $\frac{1}{\sqrt{2}}\left[\begin{array}{cc}1 & 1 \\ 1 & -1\end{array}\right]$ which transforms the state $|0\rangle$ to $\frac{|0\rangle+|1\rangle}{\sqrt{2}},|1\rangle$ to $\frac{|0\rangle-|1\rangle}{\sqrt{2}}$. In general, %one qubit gate that works on a single qubit can be represented by a $2 \times 2$ matrix, and 
one gate that works on $n$ qubits can be represented by a $2^n \times 2^n$ matrix.
Fig.\ref{fig:circuit_exmple} shows an example of a 4-qubit variational quantum eigensolver (VQE) circuit. Each horizontal line describes the time evolution of the state of one qubit from left to right. 
%which dictates how the different gates are executed. 
Gates operating independently on different qubits can be executed simultaneously, for example, the H gates operating on $q_0$ and $q_2$. Gates on the same qubit have to respect the order of the gates. For example, the CNOT gate (denoted as $\oplus$) operating on $q_0$ and $q_1$ must wait until the H gate on $q_0$ and the CNOT gate on $q_1$ and $q_2$ are finished. As another important term, 
a \textit{front layer} is defined by the set of all gates that have no unexecuted predecessors in the circuit. These gates can be executed instantly and concurrently from a software perspective. As shown in Fig. \ref{fig:circuit_exmple}, the first three H gates on $q_0, q_2, q_3$ are the front layer. 

%\vspace{-1.5ex}
%\subsection{Distributed Quantum Computing (DQC)}
%\vspace{-1ex}
%Recent advances in quantum networks support 
\textbf{Distributed quantum computing (DQC).}
DQC uses multiple quantum processing units (QPUs) to complete a large quantum computing job jointly via a quantum network\cite{pompili2021realization,hermans2022qubit,niu2023low,magnard2020microwave,li2024high}. %Similar to classical distributed computing, distributed quantum computing relies on remote communications. 
When a gate is operated by two qubits on different QPUs, called a remote gate, 
DQC uses remote communication to establish an EPR pair between the QPUs.

\vspace{-1ex}
\section{Quantum Cloud Model}
\vspace{-1ex}

\begin{figure}[t]
\setlength{\abovecaptionskip}{0.cm}
\centerline{\includegraphics[width=0.39\textwidth]{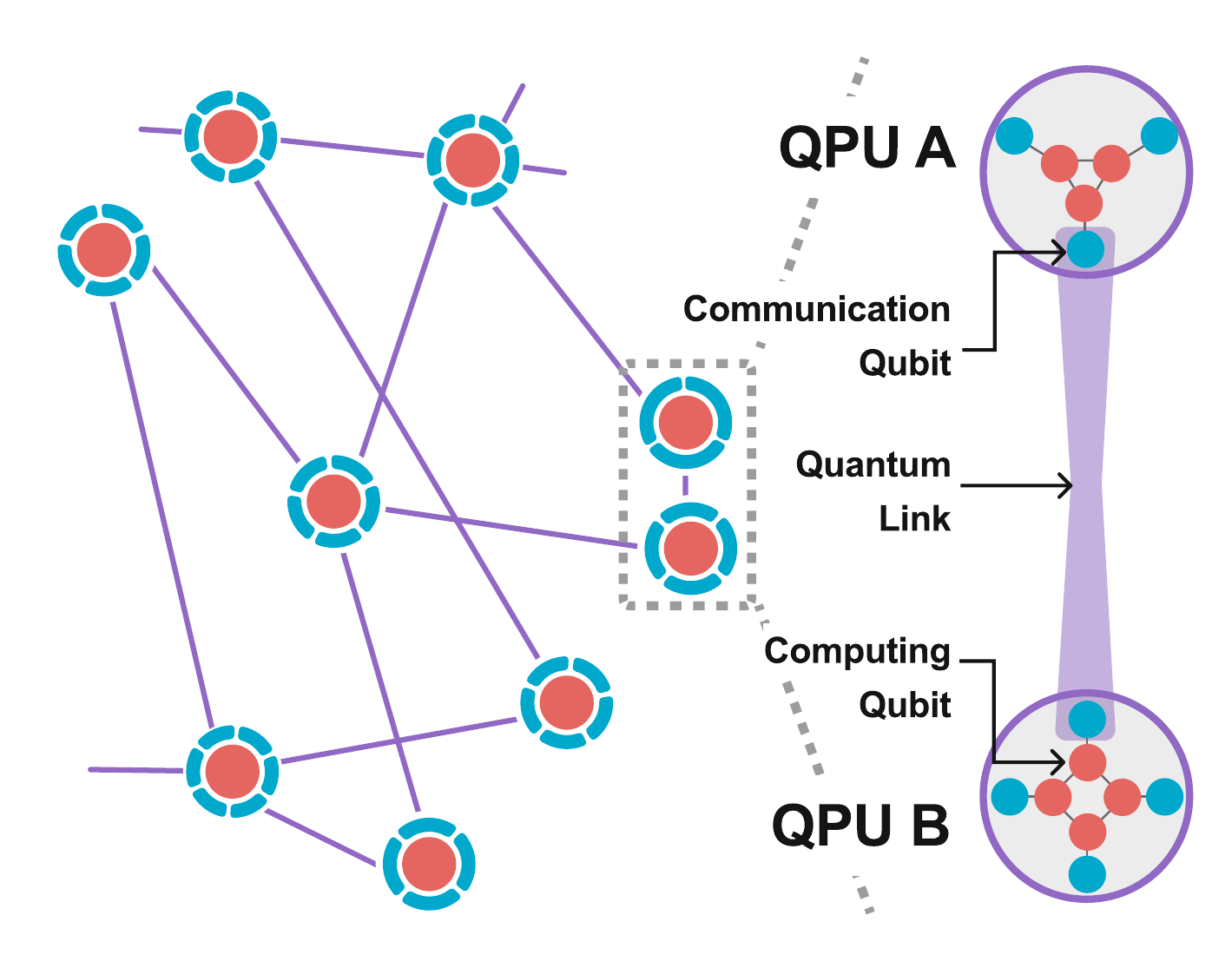}}
\vspace{-1.5ex}
\caption{Quantum Cloud}
\label{fig:cloud_model}
\end{figure}

% This paper presents QCloud, a novel quantum cloud architecture that includes multiple QPUs inter-connected by quantum switch as shown in Fig. \ref{cloud_model}. Different from previous work on designing large-scale distributed quantum computers with single quantum switches, we adopt a strategy with multiple switches. This design is inspired by previous work on designing traditional data centers\cite {guo2009bcube} which includes multiple mini-switches to speed up bandwidth-intensive applications. With this design, we can delegate high-cost communication operations in distributed quantum computing to this switch and delegate 2-qubit gates to two neighboring QPUs and thus improve both the efficiency and performance of processing a quantum circuit in the cloud setting. 
This section introduces the system model for the quantum cloud discussed in this work, an overview of our model is shown in Fig.~\ref{fig:cloud_model}. 
%An example of the quantum cloud model is shown in Fig.~\ref{fig:cloud_model}, where a network of QPUs are inter-connected by quantum links. %, forming a network of QPUs. 
%, whose components include QPUs. 
%As shown in Fig. \ref{cloud_model}, the quantum cloud model in our work mainly includes multiple inter-connected QPUs, where QPUs contain computing qubits and communication qubits.
% 别忘了加一些ionq architecture 的引用

\textbf{QPU model.} Each QPU 
%Quantum processor serves as the basic computing unit in the quantum cloud; each quantum processor 
is also equipped with a classical computer, which is used to manipulate and manage the quantum processor and transmit classical information such as measurement results to other QPUs and the controller. Each QPU includes two types of qubits: computing qubits to perform quantum gates and communication qubits that assist computing qubits for remote gates. 
%Much experiment and theoretical progress have been made in recent years about the solution for interconnecting different quantum processors to address the scaling problem~\cite{monroe2013scaling,gonzalez2021scaling,chow2021ibm}.  Fig. \ref{qpu_model} gives an example of a 4-qubit QPU; in our work, we assume an all-connected topology: Each computing qubit is connected, which means a 2-qubit-gate can be applied on any two qubits directly. Besides computing qubits, each QPU is equipped with several communication qubits to generate EPR pairs with neighboring QPUs. In our work, we also assume that the communication qubit is directly linked to each local computing qubit. 

\textbf{Controller.}
The main responsibility of the quantum cloud controller is to find the placement for each submitted circuit, and after that, it needs to decide resource allocation for all currently placed circuits to complete their execution. It also monitors the status of each QPU, such as the available computing and communication qubits. 

%\subsection{Quantum Links and Network Topology}
% In our work, we model quantum link as a channel to establish entanglement between QPUs and between QPU and quantum switch. In experiment, quantum link can be realized by optical fiber to transfer flying photonic qubit and a series of hardware such beam splitters and photon detectors to perform. With flying photons sent to a intermidate node, a series of operations will be performed with these hardware and will eventually result in two qubits entangled in adjacent QPUs. Since  ....... is hard, and in this work we only assume a limited number quantum links exist in a QPU. 

% Scaling this entangling scheme to multiple nodes requires each elementary link to be phase-stabilized independently (Fig. 2B), posing a number of new challenges. The different links, and even different segments of the same link, will generally be subject to diverse noise levels and spectra. 
\textbf{Quantum links and network topology.}
In our model, a quantum link is a channel for establishing entanglement between QPUs. %Implemented using optical fibers, 
These links carry flying photonic qubits, with devices such as beam splitters and photon detectors to realize the entanglement process. %To be noticed in this framework, 
Each QPU is assumed to have a limited number of outgoing quantum links. %This limitation is based on the fact it's hard to multiplex a quantum link for multiple nodes~\cite{pompili2021realization}, and it also needs to consider the need to separate communication qubits from computing qubits~\cite{monroe2014large}. 
Hence each QPU is connected to several (but not all) other QPUs by quantum links, forming a fixed topology of the QPU network. 
%as shown in Fig.~\ref{fig:cloud_model}. 

% is not a reflection of the broader network's capacity but rather a constraint on the individual QPU's ability to establish direct entanglement connections.  Our approach thus emphasizes the strategic deploy
%\subsection{Model for Quantum Gate and Remote Gate}
%Based on our introduction to distributed quantum computing and quantum gate, we can see that 
\textbf{Models for local gates and remote gates.}
In DQC, the latency of a remote gate includes the time for EPR preparation, local gate, and measurement. To model the latency of these operations, we use the measurement results of the latency of different quantum operations from IBM's quantum platform~\cite{ibmquantumcloud} and recent experiments~\cite{pompili2021realization}. As shown in Table \ref{tab:operations_latency}, We can see a remote gate consumes much longer time than a local gate. In Table~\ref{tab:operations_latency}, we compare different operations to the execution time of one CX gate.
Besides long execution time, another property of EPR pair generation is that its success is probabilistic. A failed EPR generation also consumes communication qubits. 
%Thus when performing a remote gate, the success of it is also probabilistic. More specifically, we model each success rate of each EPR pair generation by $p$, and each EPR pair will consume one communication qubit on each QPU. Suppose the number of EPR pair trials is $n$, the the success probability of one remote gate can be computed by $$P = 1- (1-p)^n $$
\vspace{-2ex}
\begin{table}[t]
\centering
\begin{tabular}{|l|c|}
\hline
\textbf{Operation} & \textbf{Latency} \\
\hline
Single-qubit gates & $t_{1q} \sim 0.1 \, \mathrm{CX}$ \\
\hline
$\mathrm{CX}$ and $\mathrm{CZ}$ gates & $t_{2q} = 1 \, \mathrm{CX}$ \\
\hline
Measure & $t_{ms} \sim 5 \, \mathrm{CX}$ \\
\hline
EPR preparation & $t_{iep} \sim 10 \, \mathrm{CX}$ \\
\hline
\end{tabular}
\caption{Summary of operations and latency}
\vspace{-6ex}
\label{tab:operations_latency}
\end{table}

\
% 先写一段介绍在quantum cloud情况下circuit的model sub-circuit with communication

\section{Problem Formulation}
\vspace{-1ex}
%In this section, we first introduce the motivation and challenges of our problem, followed by the formulation of two sub-problems that enable multi-tenancy in the quantum cloud. 
To execute multiple quantum circuits concurrently in a cloud setting, we need to address the following questions:

\begin{enumerate}
    \item Which circuits should we place if the current resource cannot support all circuits? Given a quantum circuit, which QPUs should the quantum cloud select to execute the circuit? %With a set of selected QPUs, which qubits should be assigned to which QPU?
    \item After placing the circuits onto QPUs, how do we decide the strategy for communication resource allocation?
\end{enumerate}
These two sub-problems are the main focus of this work. We define the first problem as \textbf{\textit{multi-tenant circuit placement}} and the second problem as \textbf{\textit{quantum network scheduling}}. 

\subsection{Design Objectives}
%Determining the optimal placement of quantum circuits and making resource allocation decisions within a batch in a multi-tenant setting is challenging due to the need to consider various factors from both the quantum cloud's and the user's perspectives. Their targets might often conflict. Here, 
We outline some key considerations of achieving optimal placement of quantum circuits and making resource allocation decisions.%, which will later be incorporated into the problem formulation: 

\begin{enumerate}
    \item \textbf{Minimizing the network cost}: 
    %Different from the local quantum gate on a single QPU, the 
    Remote gates that depend on EPR pairs are much more expensive and time-consuming than local gates. When partitioning a quantum circuit, we need to ensure the sum of the cuts is minimized so that the expensive remote communication will also be minimized. 

    \item \textbf{Dynamics in quantum cloud}: The resource availability in a quantum cloud changes dynamically with incoming jobs.
    %, and the number of available computing qubits on each QPU also changes. 
    Thus, we need to enforce the capability constraint.
    %, ensuring that the number of qubits on each partition is smaller than the number of computing qubits on the QPU that the partition is mapped to. 
    Moreover, we cannot focus solely on optimizing the allocation of each single circuit, which might lead to hard allocation for other circuits in the queue. %A simple strategy can be adopted: allocate a quantum circuit to a tightly connected component on the quantum cloud and leave the rest of the QPUs tightly connected.
    \item \textbf{Minimizing job completion time and maximizing quantum resource utilization}: Reducing job completion time is essential for improving user experience. This objective can also be reflected in improving resource utilization and avoid wasting qubits. 
    %When distributing circuits to the quantum cloud, efficient allocation is crucial to avoid wasting qubits and to ensure optimal use of available resources.
    %\item \textbf{}: For each job submitted to the quantum cloud, 
    %By minimizing completion times, we can improve the overall system throughput, enhance user experience, and boost the performance of quantum circuits.
\end{enumerate}
Our main design aims to achieve these goals. Specifically, during the multi-tenant circuit placement step, we strive to minimize the number of remote operations while maximizing quantum service utilization to avoid wasting qubits. In the quantum network scheduling step, we aim to further improve job completion time by effectively capturing the structure of the quantum circuits. With these objectives in mind, we present the formulation of our two problems.

\vspace{-1ex}
\subsection{The Multi-tenant Circuit Placement Problem}
\vspace{-2ex}

The model of the quantum cloud is depicted by $G=(V, E)$, where $V$ represents the collection of QPUs and $E$ denotes the set of connections between them, with an edge $e$ existing between any two QPUs connected by a quantum link. 
Given a batch of quantum circuits $T_k = \{T_1, T_2, \ldots, T_N\}$, we use a binary variable $x_k$ to denote whether the $k$-th circuit will be chosen to find placement in the current decision round. We use $D_{ij}^{k}$ to denote the number of 2-qubit gates between $q_i^{k}$ and $q_j^{k}$. We use $C_{ij}$ to denote the communication cost of remote gate operations between QPU $i$ and QPU $j$. $C_{ij}$ can be defined in many ways since executing one remote gate may require multiple EPR pair generation attempts. It also depends on the distance between two QPUs since it may require entanglement swapping at intermediate nodes. For illustration, in our circuit placement step, we define $C_{ij}$ as the length of the path between QPU $i$ and QPU $j$. Define \( \pi: [1, \ldots, N] \rightarrow [1, \ldots, |V|] \), where \( \pi(q_i^{k}) \) indicates which QPU the $i$-th qubit of the $k$-th circuit will be mapped to. Also, $\text{Rem}(V_i)$ is the remaining qubits on $V_i$. We formulate the multi-tenant circuit placement as follows:

\begin{align}
    \min & \sum_{k=1}^{N} \sum_{i=1}^n \sum_{j=1}^n  x_{k}D_{ij}^{k} C_{\pi(q_i^{k}) \pi(q_j^{k})} \label{eq:min1} \\
    \min & \sum_{i=1}^n \text{Rem}(V_i) \label{eq:min2} \\
    \text{s.t.} \quad 
    & \sum_{i : \pi(i) = j} |Q_i| \leq \text{Capacity}(V_j), \quad \forall j \in \{1, 2, \ldots, n\} \label{eq:capacity} \\
    & x_{k} \in \{0,1\}, \quad \forall k \label{eq:binary} \\ 
    &\sum_{k=1}^N x_k \geq 1 \label{eq:conss}\\ 
    & R(V_j) \leq \epsilon, \quad \forall j \in \{1, 2, \ldots, n\} \label{eq:threshold}
\end{align}
\vspace{-1ex}
where    
\vspace{-1ex}
\begin{equation}
\begin{aligned}
    R(V) &= \sum_{k=1}^{N} \sum_{i=1}^n \sum_{j=1}^n D_{ij}^{k} \cdot \delta(\pi(q_i^k), \pi(q_j^k), V_j), \\
    &\quad \forall j \in \{1, 2, \ldots, n\} \label{eq:Rvj}
\end{aligned}
\end{equation}

and \( \delta(\pi(q_i^k), \pi(q_j^k), V_j) \) is defined as:

\[
\delta(\pi(q_i^k), \pi(q_j^k), V_j) = 
\begin{cases} 
1 & \begin{aligned}
    \text{if } &\pi(q_i^k) \neq \pi(q_j^k) \\
    &\text{and } (\pi(q_i^k) = j \text{ or } \pi(q_j^k) = j)
\end{aligned} \\
0 & \text{otherwise}
\end{cases}
\]

\noindent
The function \( \delta(\pi(q_i^k), \pi(q_j^k), V_j) \) is an indicator function that returns 1 if qubits $q_i^k$ and $q_j^k$ are mapped to different QPUs and one of them is mapped to $V_j$, and 0 otherwise. The term \( R(V_j) \) represents the number of remote operations involving QPU $V_j$. It is calculated as the sum of all 2-qubit gates between qubits mapped to different QPUs, where one of the QPUs is $V_j$. Objective~\ref{eq:min1} minimizes the total communication cost for all circuits. Objective~\ref{eq:min2} minimizes the sum of unused computing qubits among all QPUs to maximize resource utilization. Inequation~\ref{eq:capacity} ensures that the sum of used qubits on each QPU does not exceed its capacity. \ref{eq:binary} ensures that $x_k$ is a binary variable. Inequation~\ref{eq:conss} ensures at least one circuit must be selected which avoids the case that all $x_k =0 $. Inequation~\ref{eq:threshold} enforces that the number of remote operations involving each QPU does not exceed a threshold $\epsilon$. 

%for INFOCOM
%\textbf{Discussion}: In our problem formulation, we aim to maximize qubit utilization and minimize the total cost of all circuits rather than simply maximizing the number of circuits to be scheduled. This is because merely maximizing the scheduled circuits may not improve overall system throughput and circuit performance. Maximizing the number of scheduled circuits can lead to multiple circuits competing for communication resources on a single QPU, resulting in congestion. To mitigate this issue, we include Eq.~\ref{eq:threshold} to ensure that when multiple circuits overlap on a single QPU, they do not incur an excessively large number of remote operations and the total number of remote operations should be smaller than a threshold. This constraint helps to prevent congestion and maintain efficient operation across the quantum cloud.

% \begin{equation}
% \min \max_{u \in \mathcal{N}} c_u
% \end{equation}
% \begin{align}
% \text{s.t.}\sum_{u \in \mathcal{N}_i} x_u &\leq M_i \quad \forall t \label{eq1} \\
% s_{u_j} &\geq c_{u_i} \quad \forall (u_i, u_j) \in \mathcal{A} \label{eq2} \\
% c_u &= s_u + d_u(x_u) \label{eq3} 
% % x_u &\geq 0, \quad s_u \geq 0 \quad \forall u \in \mathcal{N} \label{eq4}
% \end{align}
 % For each identified partition, the aim is to accurately map it to a corresponding QPU $V_i$ within the set $V$, ensuring efficient execution of the job across the quantum cloud infrastructure. 

\begin{figure*}[t]
\setlength{\abovecaptionskip}{0.cm}
\centerline{\includegraphics[width=1.0\textwidth]{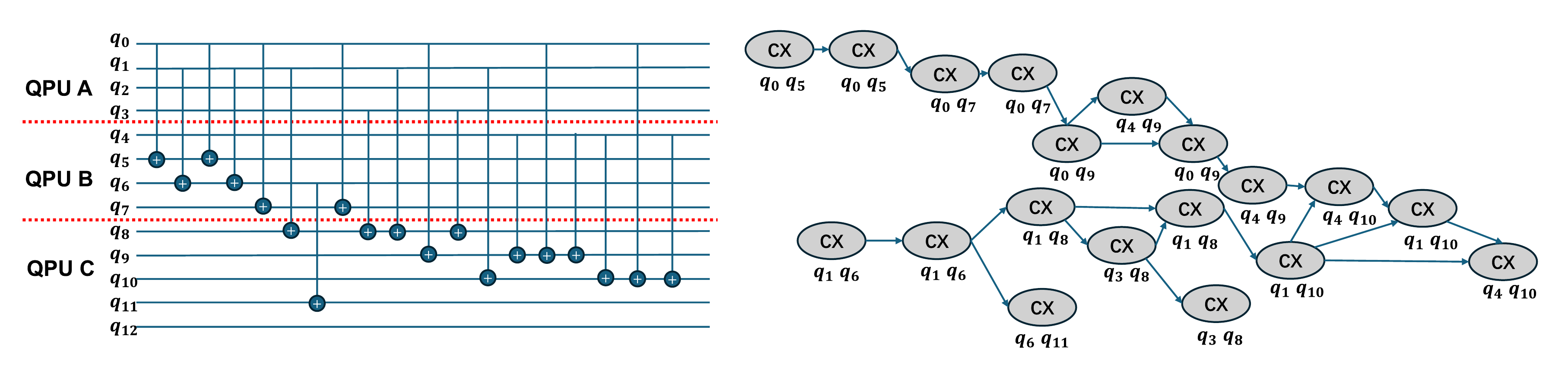}}

\caption{(a) An example circuit spams three QPUs. (b)Corresponding remote DAG that only contains inter-QPU remote gates}
\label{example_dag}
\vspace{-1ex}
\end{figure*}
%\vspace{-1ex}
\textbf{Complexity analysis and NP-hardness.}
% current version reduce this to QAP 
%Here, we show our analysis of the complexity of circuit placement in our quantum cloud setting. 
We first consider the simplest setting of our problem, where we only find the placement of a single circuit with only compatibility constraints. Given $C_{ij}$ and $D_{ij}$ are real values, the single circuit placement problem we consider here falls into the Quadratic Assignment Problem (QAP)~\cite{loiola2007survey}, which is known to be the most difficult problem in NP-hard class~\cite{sahni1976p}. The reduction can be constructed easily with the following: Given one instance of the QAP problem with facilities $F$ and locations $L$, a weight function $w: F \times F \longrightarrow R$, and a distance function $d: L \times L \longrightarrow R$. The corresponding circuit placement problem can be constructed by: $Q=F$ , $E=L$, $D_{ij} = W_{ij}$ and $d_{ij}=C_{ij}$. And by this construction, we can get an equivalent QAP problem. 

To extend this problem to the multi-tenant setting, consider multiple instances of the QAP, one for each circuit. For each circuit $k$, we create qubits corresponding to the facilities in the QAP instance and treat each circuit as an independent QAP. We then define the communication costs $C_{ij}$ to model interactions between qubits from different circuits. Then, we can formulate the multi-circuit problem as one  Multi-objective Linear Programming problem. Due to the page limitation, we omit the details here. This reduction shows that the multi-tenant circuit placement problem is at least as hard as multiple instances of the QAP. 
%Since solving the multi-tenant problem would solve each instance of the QAP, and QAP is NP-hard, it follows that our multi-tenant quantum circuit placement problem is also NP-hard. 
It easily gives that the multi-tenant quantum circuit placement problem is also NP-hard.

\subsection{Network Scheduling}

 % In the Flow Scheduling step, we will determine how to allocate communication resources(communication qubits) to each remote operation. After job scheduling, we get the mapping of a group to QPUs and we can identify which gates on the original circuit become remote gates. By keeping the order of the original DAG and ignoring local 2-qubit and 1-qubit gates, we can get a modified dag that only keeps remote 2-qubit gates. As shown in Fig. \ref{flow_scheduler}, (a) shows an example 6-qubit gate; for convenience, we only show 2 qubit gate and ignore the directions. Suppose after job scheduling, $q_0$ is mapped to QPU A and $q_1$, $q_2$ and $q_3$ are mapped to QPU B and $q_4$ and $q_5$ are mapped to QPU C. After this mapping, 2-qubit gates M and N are ignored on Fig .\ref{flow_scheduler} since they become local gates. Performing remote gate operations first requires EPR pair generation. However, \textbf{Communication qubits on each QPU is limited, and generating EPR pairs is in nature a probabilistic operation. To perform a remote gate, we have to make sure at least one EPR pair between communication qubits on two QPU is ready. Thus, we need to determine how many pairs of communication qubits we need to prepare to perform EPR pairs generation for each remote gate on the remote DAG} 
\textbf{Network scheduling example.}
Fig.~\ref{example_dag} (a) shows all remote gates of a  13-qubit multiplier circuit, %Here, we ignore single-qubit gates and only keep two-qubit inter-QPU remote gates, 
and (b) shows the corresponding DAG. To be noticed, we ignore single-qubit gates and only keep two-qubit inter-QPU remote gates, we  With this example, we can see various contentions on communication qubits exist. For example, the first CX gate between $q_0$ and $q_5$ and the second  CX gate between $q_1$ and $q_6$: These two gate spans the same set of QPUs (QPU A and QPU B) and will both rely on communication qubits on two QPUs. Also, the CX gate between $q_6$ and $q_{12}$ and the CX gate between $q_0$ and $q_7$, both gates will rely on the communication qubits on QPU B.
However, \textbf{communication qubits on each QPU is limited, and generating EPR pairs is a probabilistic operation. To perform a remote gate, we must ensure that at least one EPR pair between communication qubits on two QPU is ready. Thus, we need to determine how many pairs of communication qubits we need to prepare to perform EPR pairs generation for each remote gate on the remote DAG.} 

\subsubsection{Problem Formulation}

The network scheduling problem in our quantum setting can be formulated as follows. Let $M_i$ denote the number of communication qubits on QPU $i$.  After the circuit placement step, each job will be mapped to a set of QPUs by a function: $f: J \rightarrow \mathcal{P}(Q)$ where $\mathcal{P}(Q)$ denotes the power set of QPUs. 
%After ignoring the 1-qubit gate, and local 2-qubit gate, and keeping the logical dependency of inter-QPU 2 qubit gate as introduced earlier: 
For each job $J_i$, we can get a DAG $\mathcal{G}=(\mathcal{N}, \mathcal{A})$ where each node $u \in \mathcal{N}$ represents a remote operation between two machines $Q_j$, $Q_k$, where $j$ and $k$ depends on previous circuit placement step. Let $x_u$ be the number of communication qubits allocated to remote operation $u$. Let $c_u$ denote the makespan (total completion time) of node $u$. Then, the objective is to minimize the makespan of all remote operations while respecting the logical dependencies of DAG $\mathcal{G}$ and communication resources constraint on each QPU. The success probability of each remote operation $u$ depends on the allocated resources $x_u$, denoted by $p_u(x_u)$.
% \begin{equation}
% \min \max_{u \in \mathcal{N}} c_u
% \end{equation}
% \vspace{-2ex}
% \begin{align}
% \text{s.t.}\sum_{u \in \mathcal{N}_i} x_u &\leq M_i \quad \forall t \label{eq1} \\
% s_{u_j} &\geq c_{u_i} \quad \forall (u_i, u_j) \in \mathcal{A} \label{eq2} \\
% c_u &= s_u + d_u(x_u) \label{eq3} 
% % x_u &\geq 0, \quad s_u \geq 0 \quad \forall u \in \mathcal{N} \label{eq4}
% \end{align}
\begin{align}
\min \; & \max_{u \in \mathcal{N}} c_u \nonumber \\[6pt]
\text{s.t.}\;&\sum_{u \in \mathcal{N}_i} x_u \leq M_i && \forall i \label{eq1}\\[3pt]
&s_{u_j} \geq c_{u_i} && \forall (u_i, u_j) \in \mathcal{A} \label{eq2}\\[3pt]
&c_u = s_u + d_u(x_u) && \forall u \in \mathcal{N} \label{eq3}
\end{align}
% \begin{equation}
% \begin{aligned}
% \min \; & \max_{u \in \mathcal{N}} c_u \\[6pt]
% \text{s.t.}\;&\sum_{u \in \mathcal{N}_i} x_u \leq M_i, && \forall i \\[3pt]
% &s_{u_j} \geq c_{u_i}, && \forall (u_i, u_j) \in \mathcal{A} \\[3pt]
% &c_u = s_u + d_u(x_u), && \forall u \in \mathcal{N}
% \end{aligned}
% \end{equation}
Inequation~\ref{eq1} denotes that, for each QPU $i$,  the total allocated resources at any time do not exceed the available communication qubits on each machine $i$. Here, $\mathcal{N}_i$ is the set of operations involving machine $i$. Inequation~\ref{eq2} ensures that an operation $u_j$ can only start after its predecessor $u_i$ has successfully completed, respecting the logical dependencies in the DAG. Eq.~\ref{eq3} denotes that the completion time of an operation $u$ is the start time plus the duration, which depends on the allocated resources. It is important to note that when allocating communication qubits to each remote gate that spans $Q_i$ and $Q_j$, the corresponding resources on both QPUs need to decrease by the allocated amount to reflect the consumption of communication resources on each involved QPU.

%To summarize, we formulate the network scheduling problem in the quantum cloud with one circuit as one DAG scheduling problem, where each node on the DAG is one probabilistic event that depends on our resource allocation strategy.
% \section{QCloud} 

\section{CloudQC's Design}

\subsection{Design Overview}

\begin{figure}[t]
\setlength{\abovecaptionskip}{0.cm}
\centerline{\includegraphics[width=0.49\textwidth]{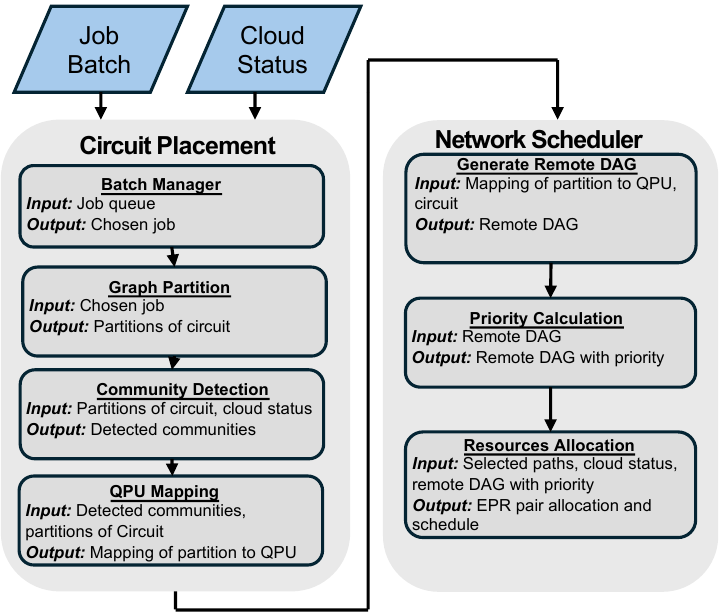}}

\caption{Overview of our scheduler workflow}

\label{workflow}

\end{figure}

\begin{figure}[t]
\setlength{\abovecaptionskip}{0.cm}
\centerline{\includegraphics[width=0.49\textwidth]{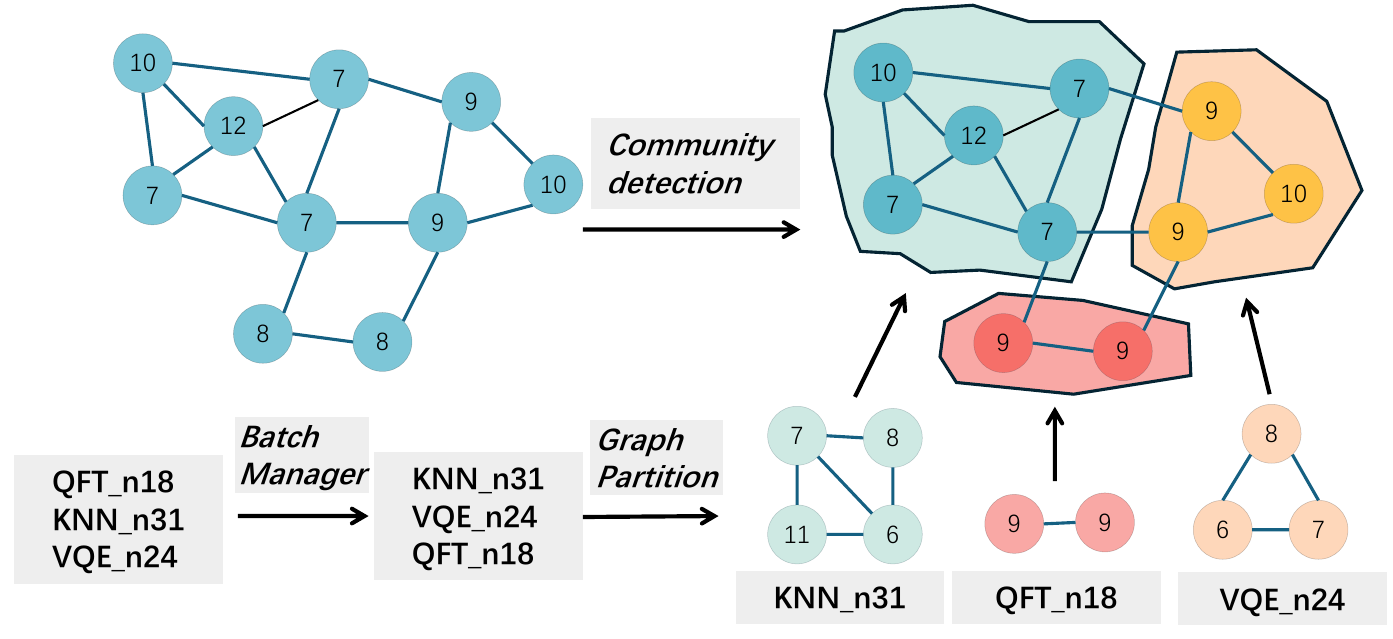}}

\caption{Example of a quantum cloud with three DQC jobs.}

\label{fig:community}
\end{figure}
\vspace{-1ex}

We present the workflow of CloudQC in Fig.~\ref{workflow}, which encompasses two main steps: circuit placement and network scheduling. 
CloudQC first determines the optimal placement for a batch of circuits, which involves circuit partitioning and assigning each partition to a suitable QPU to reduce remote communication costs. This starts with determining several partitioning strategies for a quantum circuit, followed by a feasibility assessment to select a compatible set of QPUs with the proposed community-detection method which will be introduced in detail later. The optimal placement is then selected through a scoring methodology. Upon finalizing the partitioning approach and corresponding QPUs, the network scheduler compiles the job and allocates the communication resource to each remote operation. Leveraging the optimal placement, CloudQC then calculates the network resource allocation for the jobs based on the priorities we defined. This includes determining the number of EPR pair generation attempts required for each remote gate. 

\subsection{Circuit Placement}

%In this section, we introduce our circuit placement method. 
Fig.~\ref{fig:community} illustrates an example of the circuit placement algorithm. Three circuits are submitted to the quantum cloud. First, we determine the processing order of the jobs using the batch manager discussed below. Next, for each job, CloudQC determines how to partition the quantum circuit using a graph-partitioning algorithm. 
%For each partitioning strategy, we apply
CloudQC then applies a modularity-based community detection algorithm~\cite{newman2006modularity} to identify a set of QPUs capable of running the job. We then map each partition to a QPU. 
%In our work, we define a placement by how the circuit is partitioned and how each partition is mapped to QPUs. Consequently, f
For each circuit, we identify several possible placements. We use a scoring-and-filtering method to evaluate the quality of each placement and select the best one.  %in the following sections.

\begin{algorithm}
\caption{Circuit Placement}
\label{alg:job_scheduler}

\KwIn{Job $J$, imbalance factor list $\alpha = \{\alpha_1, \alpha_2, \ldots, \alpha_n\}$}
\KwOut{Best Placement for $J$}

$InteractionGraph, DAG \leftarrow \text{Preprocessing}(J.\text{circuit})$\\

\eIf{$J.\text{circuit}.\text{size} < \min(\{QPU.\text{available} \mid QPU \in \text{Cloud}\})$}{
    \text{Allocate $J$ to the QPU with sufficient available resources}\\
}{
    Initialize an empty list for potential placements: $placement\_list \leftarrow \emptyset$\\
    \ForEach{$\alpha_i$ in $\alpha$}{
        \For{$i$ in range($\lceil \frac{J.\text{circuit}.\text{size}}{\text{QPU}.\text{num}} \rceil$, number of QPUs in Cloud)}{
            $res \leftarrow \text{graph\_partition}(\alpha_i, i)$\\
            $placement \leftarrow \text{Find\_placement}(res)$\\
            $time \leftarrow \text{estimate\_time}(DAG, res)$\\
            $communication\_cost \leftarrow \text{calculate\_cost}(res, InteractionGraph)$\\
            $placement.\text{score} \leftarrow \text{score}(time, communication\_cost)$\\
            Add $placement$ to $placement\_list$\\
        }
    }
    $best\_placement \leftarrow \text{Find\_highest\_score}(placement\_list)$
}

\Return $best\_placement$
\vspace{-1ex}
\end{algorithm}
\vspace{-1ex}
\textbf{Batch manager.}
CloudQC works for two job processing modes: The batch mode and the incoming job mode. In the incoming job mode, jobs arrive one after another and CloudQC processes them in a first-in-first-out order.
%, where we will always process the earliest job that arrives. However, 
In the batch mode, multiple jobs arrive at the same time and CloudQC needs to decide the optimal order of executing the jobs.
%in the batch will be crucial. 
%From the quantum cloud perspective, we want to improve the utilization of quantum hardware, which means we do not want running but unused QPU and physical qubits, and we also want to improve the throughput. From the user perspective, we want to improve job completion time and performance. 
One naive approach will use a greedy approach: sorting the circuits based on a metric and allocating as many circuits as possible. However, such a method may result in partitioning one circuit into too many pieces which will result in massive communication costs and thus will eventually result in longer JCT. 

We determine the order of the circuits by this metric:

\begin{equation}
    I_i = \lambda_1 \cdot \frac{\#CNOTs}{n_i} + \lambda_2 \cdot \ n_i + \lambda_3 \cdot d_i
\end{equation}
where $n_i$ denotes the number of qubits in $i_th$ job in the circuit, and $d_i$ denotes the circuit depth of $i_th$ job. Our metric is based on the following considerations: 1) $\frac{\#CNOTs}{n_i}$ denotes how 'dense' one circuit is in terms of 2-qubit interactions. One circuit with more 2-qubit interaction will likely have more remote gates when distributed to multiple QPUs. Consequently, circuits with a higher density of 2-qubit gates are more likely to suffer from increased latency and reduced fidelity due to the overhead of managing these remote interactions. 
%Therefore, accounting for the density of 2-qubit gates in our scheduling and allocation strategies is crucial to minimize these adverse effects and ensure efficient and reliable quantum computations. 
2) $n_i$ represents the number of qubits in the job, which directly correlates to the job's resource requirements. 
%A job with more qubits will require more physical qubits on the quantum processing units (QPUs), and managing these resources efficiently is crucial for maximizing utilization. 
3) $d_i$ denotes the depth of the circuit, which reflects the execution time of the circuit. Deeper circuits will take longer to execute, and thus their impact on overall throughput and job completion time is significant. By incorporating these factors, the metric $I_i$ aims to balance the trade-offs between communication costs, resource utilization, and execution time. The weights $\lambda_1$, $\lambda_2$, and $\lambda_3$ allow for tuning the importance of each factor based on specific priorities and goals. 
%This metric guides the scheduling and allocation process in the batch model to achieve an optimal execution order that improves both hardware utilization and job performance from a quantum cloud and user perspective. With this metric, we can determine the execution order of jobs in the batch.

\textbf{Circuit placement summary.}
After determining the job we need to execute in the batch, we need to find the placement of each job. 
The objective of job placement is to find the best placement that distributes a quantum circuit to several QPUs of the quantum cloud. Our method is summarized in Algorithm.~\ref{alg:job_scheduler}. 
Our method works in a filtering-and-scoring fashion: For each circuit to be executed, we first find several methods to partition the circuit by tuning the imbalance factor of the graph partition algorithm, detailed later. Then, for each part, we first use the community detection algorithm to find a set of QPUs and apply a simple heuristic to map the partition to select QPUs. Then, for each placement we find, we evaluate it by the following scoring function: 
%\begin{equation}
$S = \alpha \times \frac{1}{T} + \beta \times \frac{1}{C}$
%\end{equation}
, where $T$ is the estimated running time of the quantum circuit, $C$ is the communication cost, and with this function, we consider both the performance and execution time of the circuit.

\textbf{Preprocessing.} CloudQC first generates a Directed Acyclic Graph (DAG) representing the logical dependency of the gates of each quantum circuit. Each single-qubit gate has to wait until its parent gate is finished, and similarly, a remote $\operatorname{CNOT}\left(q_i, q_j\right)$ gate can be executed only when all previous gates on $q_i$ and $q_j$ have finished. The interaction graph is a weighted graph where the vertices are qubits of the circuit and the edge denotes the interaction of two qubits, the weight describes how many 2-qubit gates two qubits have. As shown in Fig. \ref{example_dag}, (a) shows an example 13-qubit adder circuit, and (b) shows the corresponding DAG. 
% After preprocessing, we check whether the circuit can be scheduled directly on a QPU without distributing it over several QPUs. If such a QPU exists, we schedule the circuit on the corresponding QPU directly since executing the circuit on a single QPU directly can avoid remote communication that results in a long completion time and degraded performance. 
% Then, the scheduler needs to find one 'good' placement for one circuit, 
% Here, we used a filtering-and-scoring strategy: The job scheduler first finds several methods to distribute a quantum circuit, which involves determining which logical qubits in the quantum circuit should be grouped together.

\textbf{Partitioning quantum circuit.} When partitioning quantum circuits, there are the following considerations: 1) Minimize communication cost. A remote gate depends on the remote EPR pair and is much more expensive and time-consuming than a local 2-qubit gate. Thus, we want to minimize remote communication. 2) Reduce execution time. 
%In quantum computing, performance also decreases as execution time. 
The execution time determines the job completion time.

We first apply a graph-partition algorithm~\cite{karypis1997metis} to the interaction graph based on these considerations. The algorithm aims to divide a graph into smaller, interconnected subgraphs, with the goal of minimizing the number of edges across different subgraphs while maintaining a certain load-balancing level. This is because extremely
uneven partitions may not fit the resource availability on QPUs.
%which means that the size of each subgraph is balanced. However, when considering load-balancing, there may be a chance that the communication cost is not optimal. An example with a different imbalance factor is shown in Fig .\ref{partition}. We can see with a default setting, nodes 1,2,3 are grouped together and nodes 4,5 are grouped together with a cut 2. A larger imbalance factor results in nodes 1,2,3,4 are grouped together with a cut of 1. 
Thus, in our implementation, we also tune the imbalance factor, which defines how imbalanced the result of graph partitioning can be. 

% After the graph-partitioning process, we ca n get a modified version of DAG which only contains remote gates across multiple QPUs. The execution time of such a circuit can be estimated by the length of the longest path of modified DAG since, in our work, we assume the time of the remote gate is much longer than that of the local gate. 
\textbf{Finding feasible QPU sets.} For each quantum circuit partition result, CloudQC needs to find a set of QPUs with available resources to accommodate such partition. %Finding a set of QPUs to accommodate each partition is not trivial, and
This is a challenging task because
the choice of QPUs affects not only the performance of the current circuit but also future circuits. We need to consider the following factors in addition to resource avaiability:
%1) Capability. When mapping a part of the circuit to a QPU, we must ensure that the chosen QPU has enough computing qubits for a given partition. 2
1) Communication cost. When mapping the partition to QPUs, we need to ensure that the parts with large inter-communication costs should be placed onto QPUs with short network distances, because multi-hop communication will further increase the network cost and job failure rate. %so that remote-gate operation will not spam other QPUs, which may require entanglement swapping operation that will further introduce overhead. 
2) Future resource availability. When choosing a set of QPUs, we hope such placement can provide resource availability for future jobs. 
%As shown in Fig. \ref{flexibility}, the second placement provides more flexibility than the first placement. Since placement 2, after circuit 2 finishes, it will provide a more connected set of QPUs for future jobs since it didn't leave any 'fragment' of QPUs
% This algorithm suits our setting as it considers the quantum cloud topology, ensuring the selected QPUs are tightly interconnected, as well as the rest. Encoding the number of computing qubits and available communication qubits into the edge weight ensures the selected QPUs have abundant computing qubits. 
% To summarize, the output of our methods has the following nice features: 
One good example is that after the current placement, the remaining QPU resources are still within short network distances. 
To achieve these goals, we use a modularity-based community detection algorithm~\cite{newman2006modularity} to find feasible QPU sets. A community detection algorithm identifies groups of nodes in a network that are more densely connected internally than with the rest of the network. A modularity-based community detection algorithm optimizes the modularity metric, which quantifies the quality of the division by comparing the density of links within communities to the density of links between communities. 

%This approach is suitable for our setting as it considers the quantum cloud topology, ensuring the selected QPUs are tightly interconnected. 
Additionally, we can embed the number of computing qubits into the edge weight. This ensures that the selected QPUs have both strong connectivity and abundant computing qubits, capturing the dynamics of the quantum cloud and reflecting the capability of the selected QPUs. 
%Although we did not consider the impact of imperfect EPRs in this work, 
The community detection method provides a powerful framework for profiling each QPU. This profiling can help measure the performance of QPUs in various aspects. For example, in future modular quantum computer designs, we might consider the reliability of quantum links between QPUs and the reliability of each QPU. This information can be easily encoded into the edge weights.
% \begin{itemize}
%     \item When choosing the QPU set, we consider both the computing qubits constraint and topology information. The chosen QPUs must have enough qubits and a tightly connected topology to form a community.
%     \item Our community detection-based method ensures that the chosen QPUs are tightly connected and the rest of the QPUs are also tightly connected, which will prevent the existence of fragments of QPUs, as shown in the example
%     \item Our community detection-based method provides the power of profiling each QPU, which may help measure the performance of QPUs in different aspects. For example, in the future modular quantum computer design, we may need to consider the reliability of quantum links between QPU and the reliability of each single QPU, and we can easily encode this information into the weight of edges. 
% \end{itemize}

\textbf{Mapping partitioned circuits to selected QPUs.} 
%After partitioning the circuit and finding QPUs, we need to find a mapping from the partition to selected QPUs. We summarized the details of our simple heuristics in the Algorithm.~\ref{overview}. 
We use a heuristic algorithm to map the partitioned circuits to QPUs. 
We calculate the graph center of the found community and the interaction graphs, which minimizes the longest topological distance to all other nodes. Then, we map the center of the remote interaction graph to the center of the detected community graph. The rest of the mapping will be expanded from the node with the highest weight edge. We perform a breadth-first search around the first logical center to place each qubit of the circuit to an available QPU with the least distance to the center. The details of the heuristic are shown in the Algorithm.~\ref{overview}. \textbf{This simple heuristic ensures that two partitions with a high communication cost will be mapped to two close QPUs in the cloud.}

\begin{algorithm}
\label{overview}
\caption{Find Placement}
\KwIn{Partition $\mathcal{P} = \{P_1, P_2, \ldots, P_n\}$, Quantum Cloud QPU Topology Graph $G_c$, Remote Partition Interaction Graph $G_p$}
\KwOut{Partition to QPU Mapping:mapping[]}
Initialize mapping[] = -1

C $\leftarrow$ CommunityDetection($G_c$)

$C_c$ $\leftarrow$ GraphCenter(C)

$C_p$ $\leftarrow$ GraphCenter($G_p$)

$mapping[C_p] \leftarrow C_c$  

$q$ $\leftarrow$ BfsQueueGen($G_c$)

\While{$q$ is not empty}{
$q_i \leftarrow q.front()$ \\
$C_c$ $\leftarrow$ GraphCenter(C) \\
\If{mapping[$q_i$] = -1}{
$mapping[q_i] \leftarrow C_c$  
}
$N(q_i) \leftarrow GetNeighbors(q_i)$ \\
$N(C_c) \leftarrow GetKClosestNode(C_c)$ \\
$mapping[N(q_i)] \leftarrow N(C_c)$ \\
$q.pop()$
}
\end{algorithm}
\vspace{-1.5ex}

\subsection{Network Scheduling }

% 在ooptunities and challenges这个section，给出每一步的需要考虑的问题，某些问题需要给出一些例子,先看一些文章学一些话术
% \textbf{Topology and capability aware placement}. When mapping a circuit to quantum cloud topology, we need to consider several aspects. The first
 % \subsubsection{Motivation Example}

\begin{algorithm}
\caption{Network Scheduler}
\label{alg:flow_scheduler}

\KwIn{Remote DAG}

Compute priorities for all nodes in the DAG \\
Initialize front layer based on in-degree and out-degree \\
% Determine remote combinations from partition data \\
\While{Remote DAG is not empty}{
    Define resource availability for each partition \\
    Identify competing sets for resource allocation \\
    Allocate resources between competing sets \\
    \ForEach{node in front layers}{
        Determine resources allocation based on the priority of node \\
        \If{node is successfully executed}{
            Update status table \\
            Prepare for node deletion and successor addition \\
        }
    }
    Update front layer and DAG based on node execution \\
}

\end{algorithm}
\vspace{-2ex}

We would like to achieve the following desirable properties of network scheduling: 
%When designing our network resources scheduler, we hope our network schedulers have the following properties:
\begin{itemize}
    \item Effectiveness: More `important' sub-jobs in the remote DAG should receive more communication resources to prevent subsequent tasks from being backlogged. When multiple sub-jobs can be executed concurrently, more important jobs should receive more resources.
    \item Starvation freedom: When multiple jobs compete for communication resources on the same QPU, no job should be starved forever.
   % \item Work conserving: Resources should be fully utilized at each step of the flow scheduling process.
\end{itemize}

We use the same example in n Fig.~\ref{example_dag} to show our observations: $q_0$ and $q_5$ and the second  CX gate between $q_1$ and $q_6$: These two gates span the same set of QPUs and can be processed in parallel. However, these two gates have different 'importance' on the DAG: the first gate is on the critical path, and there are also more gates on its corresponding paths meaning that the failure of this gate may cause more gates to be backlogged. Thus, we should assign redundant resources to the first gate to increase failure tolerance.  
%and less to the second gate since the probability of successful execution of a remote gate depends on how much communication resources we allocate. 
Another example is the seventh gate between $q_6$ and $q_11$, which crosses QPUs $B$ and $C$, and the eighth gate between $q_1$ and $q_8$, which crosses QPUs $A$ and $B$. Both gates will use communication qubits on QPU $B$, but we can see the gate between $q_1$ and $q_8$ is far less important than the first gate. Thus, CloudDC will allocate more communication resources to the ninth gate than the eighth gate.

With these observations, CloudDC first quantifies the importance of each node on remote DAG with the priority we define.  We denote the priority of node $n_i$ as $p_i$ and $\mathcal{P}(n_i)$ as the set of paths connecting node $n_i$ with any remote DAG leaves. The priority $p_i$ can now be computed by: $p_i = \max_{P \in \mathcal{P}(n_i)} |P|$,
%\vspace{-2ex}
%$$
%p_i = \max_{P \in \mathcal{P}(n_i)} |P|
%\vspace{-1ex}
%$$
where $P$ represents any path from $n_i$ to a leaf node, and $|P|$ is the length of path $P$ in terms of the number of edges. This priority calculation is based on the depth of the longest path from node $n_i$ to any leaf node in the DAG. With this priority value, we can determine whether the node is on the critical path of remote DAG and how many nodes will be blocked if the execution of the node is unsuccessful.

With these observations and our definition on priority, we summarize our method for network scheduling in Algorithm.~\ref{alg:flow_scheduler}. The network scheduling algorithm first calculates the priority of each node. Then, it initializes the front layer, which is defined by the remote operations that can be processed in parallel. %In the main loop, the flow scheduler determines how many resources can be allocated(Line 4), 
It then identifies contentions between different QPU partitions and then allocates resources to them. %After determining how many resources can be allocated to each QPU partition, 
The network scheduler repeatedly checks the current front layer on whether each gate is executed successfully and updates the front layer in the graph if some gates are completed. The above process repeats until the remote DAG is empty.

% 五大组实验，
% 1. 测circuit placement的实验，对于每个单一的circuit测，x轴不同的circuit，y轴 cost： 

\section{Evaluation}

In this section, we use experimental evaluations to answer the following questions:
\begin{itemize}
    \item How does the proposed job placement method compare to existing algorithms (Sec. VI. B)?
    \item How much can network scheduling improve compared to other resource allocation strategies in terms of job completion time (VI. C)?
    \item How is CloudQC's general performance in a multi-tenant setting (VI. D)? 
\end{itemize}
\vspace{-2ex}
\subsection{Evaluation Setting}
\vspace{-1ex}
\textbf{Workloads.} We use the quantum circuits from an existing benchmark \cite{li2023qasmbench}, whose characteristics are shown in Table~\ref{tab:quantum_circuits}. 
\begin{table}[t]
\centering
\caption{Quantum circuit characteristics}
\vspace{-2ex}
\label{tab:quantum_circuits}
\begin{tabular}{c|c|c|c}
\hline
\textbf{Name} & \textbf{\# of Qubits} & \textbf{\# of 2-Qubit Gates} & \textbf{Circuit Depth} \\
\hline
ghz\_n127  & 127 & 126 & 128 \\
bv\_n70 & 70 & 36 & 40 \\
bv\_n140 & 140 & 72 & 76 \\
ising\_n34 & 34 & 66 & 16 \\
ising\_n66 & 34 & 130 & 16 \\
ising\_n98 & 98 & 194 & 16 \\
cat\_n65 & 65& 64&66\\
cat\_n130 &130 & 129 &131 \\
swap\_test\_n115 &115 & 456 &60\\
knn\_n67 &67& 264 &36\\ 
knn\_n129 & 129 & 512 & 67 \\
qugan\_n71 & 71 &418& 72\\
qugan\_n111 & 111 &658& 112\\
cc\_n64 & 64 & 64  &195\\
adder\_n64 & 64 & 455 &78\\
adder\_n118 & 118 & 845 &132\\
multiplier\_n45 & 45 & 2574 & 462\\
multiplier\_n75 & 75 & 7350 & 1300 \\
qft\_n63& 63 & 9828 & 494 \\
qft\_n160 &160 & 25440 &1270 \\
qv\_n100 & 100 & 15000 & 701\\
\hline
\end{tabular}
\end{table}

\textbf{Implementation.} Since there is no publicly available quantum cloud simulator, we developed a customized discrete-event simulator in Python. 
%We use a simple discrete event simulator(DES) to model time and events. 
For quantum circuit analyzing, we use PytKet~\cite{sivarajah2020t}. We use PyMetis, a Python version package for Metis~\cite{karypis1997metis} for graph partition. The simulation code is available to the public~\cite{cloudqc}.

\textbf{Topology setting.}
We set the default number of QPUs in a quantum cloud to be 20; each QPU is equipped with 20 computing qubits and 5 communication qubits. We use a random topology, and we set the probability of generating an edge to be 0.3. We set the success probability of generating an EPR pair to be 0.3, consistent to existing work and the experiments \cite{shi2020concurrent,pompili2021realization}. %For the graph partition, we set the default imbalance factor list to be [1,100,100]. 

\begin{table}[t]
\centering
\caption{Number of Remote Operations of single-circuit placement}
\label{tab:experiment_results}
\begin{tabular}{c|c|c|c|c|c}
\hline
\textbf{Circuit} & \textbf{SA} & \textbf{Random} & \textbf{GA} & \textbf{CdQC-BFS} & \textbf{CdQC} \\
\hline
ghz\_n127        & 145 & 161 & 90   & 10  & 8   \\
bv\_n70          & 41  & 38  & 17   & 26  & 18  \\
bv\_n140         & 96  & 98  & 54   & 101 & 53  \\
ising\_n34       & 38  & 36  & 6    & 2   & 2   \\
ising\_n66       & 100 & 110 & 36   & 6   & 8   \\
ising\_n98       & 214 & 250 & 96   & 10  & 10  \\
cat\_n65         & 52  & 44  & 20   & 5   & 3   \\
cat\_n130        & 153 & 145 & 92   & 10  & 8   \\
swap\_test\_n115 & 398 & 472 & 294  & 352 & 192 \\
knn\_n67         & 158 & 230 & 106  & 168 & 100 \\
knn\_n129        & 528 & 720 & 374  & 376 & 220 \\
qugan\_n71       & 334 & 482 & 278  & 180 & 144 \\
qugan\_n111      & 838 & 1080 & 718 & 404 & 248 \\
cc\_n64          & 45  & 44  & 44   & 46  & 44  \\
adder\_n64       & 269 & 450 & 142  & 33  & 33  \\
adder\_n118      & 748 & 1225 & 613 & 60  & 37  \\
multiplier\_n45  & 596 & 1452 & 493 & 611 & 462 \\
multiplier\_n75  & 2100 & 6809 & 2255 & 1993 & 1766 \\
qft\_n63         & 2504 & 3202 & 2368 & 3012 & 2358 \\
qft\_n160        & 12326 & 15514 & 14246 & 14814 & 11132 \\

\hline
\end{tabular}
\end{table}

\begin{figure*}[!t]
  \centering
  \begin{minipage}[t]{0.24\textwidth}
    \includegraphics[width=\textwidth]{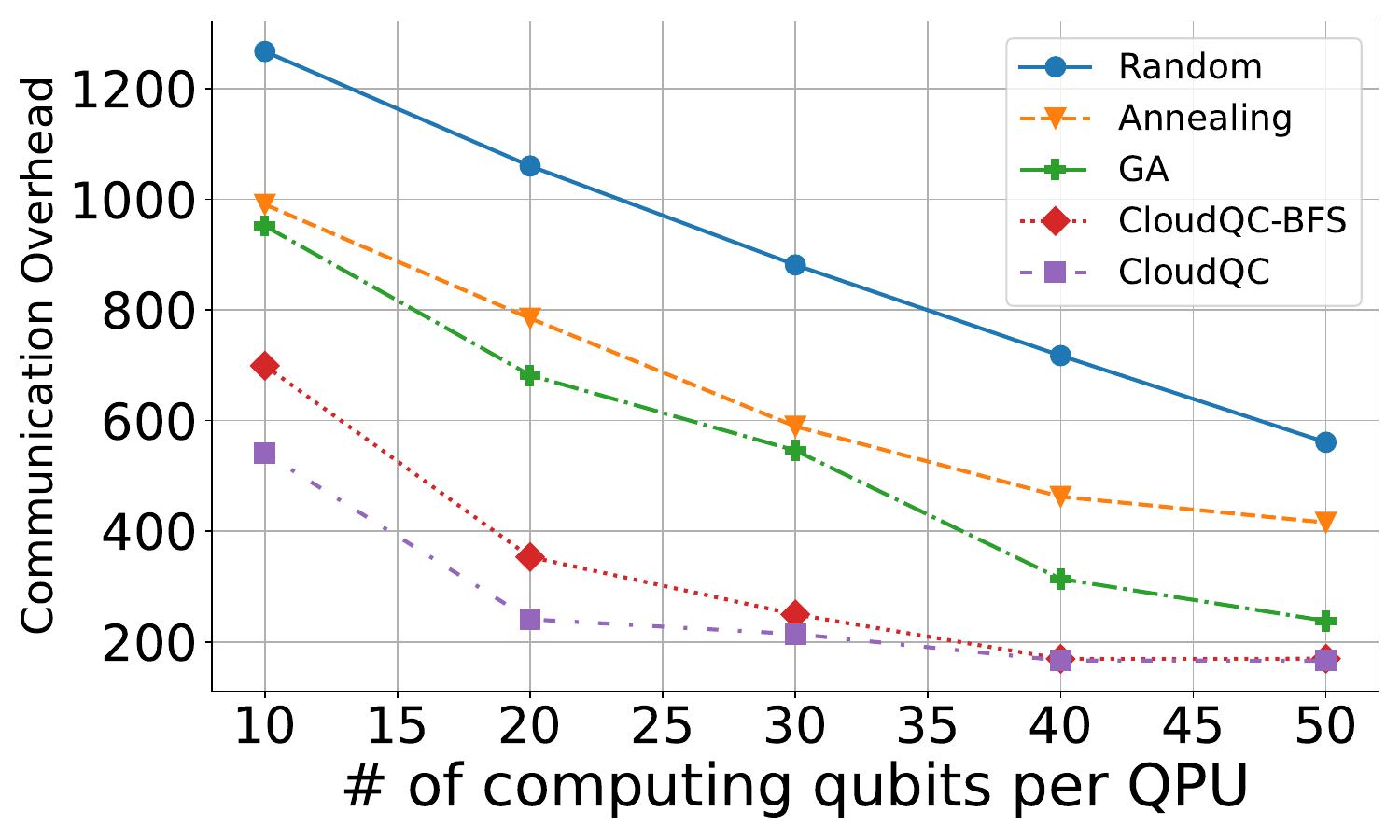}
    \vspace{-5ex}
    \caption{Overhead vs \# of computing qubits: qugan\_n111}
    \label{fig-n}
  \end{minipage}\hfill
  \begin{minipage}[t]{0.24\textwidth}
    \includegraphics[width=\textwidth]{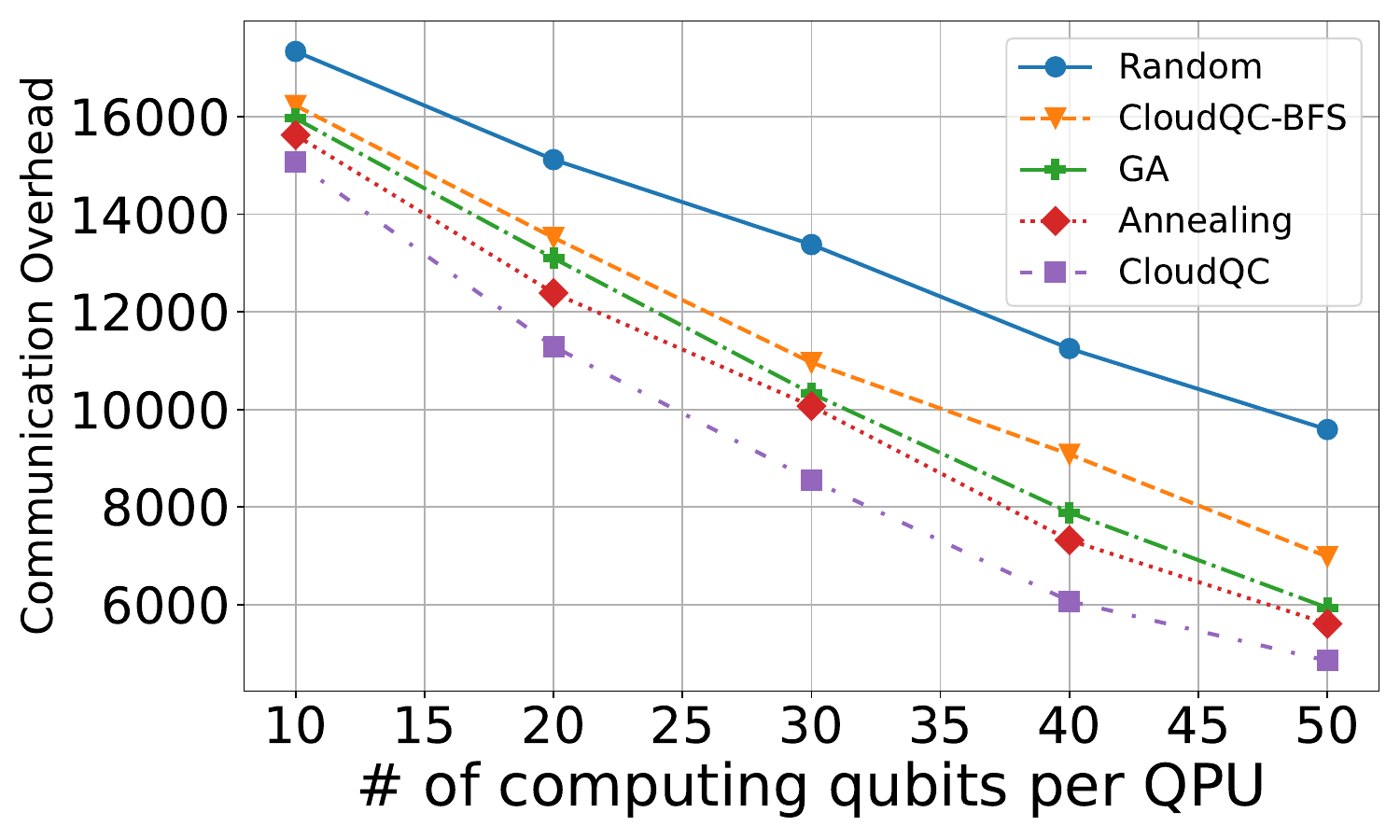}
    \vspace{-5ex}
    \caption{Overhead vs \# of computing qubits: qft\_n160}
    \label{fig-d}
  \end{minipage}\hfill
  \begin{minipage}[t]{0.24\textwidth}
    \includegraphics[width=\textwidth]{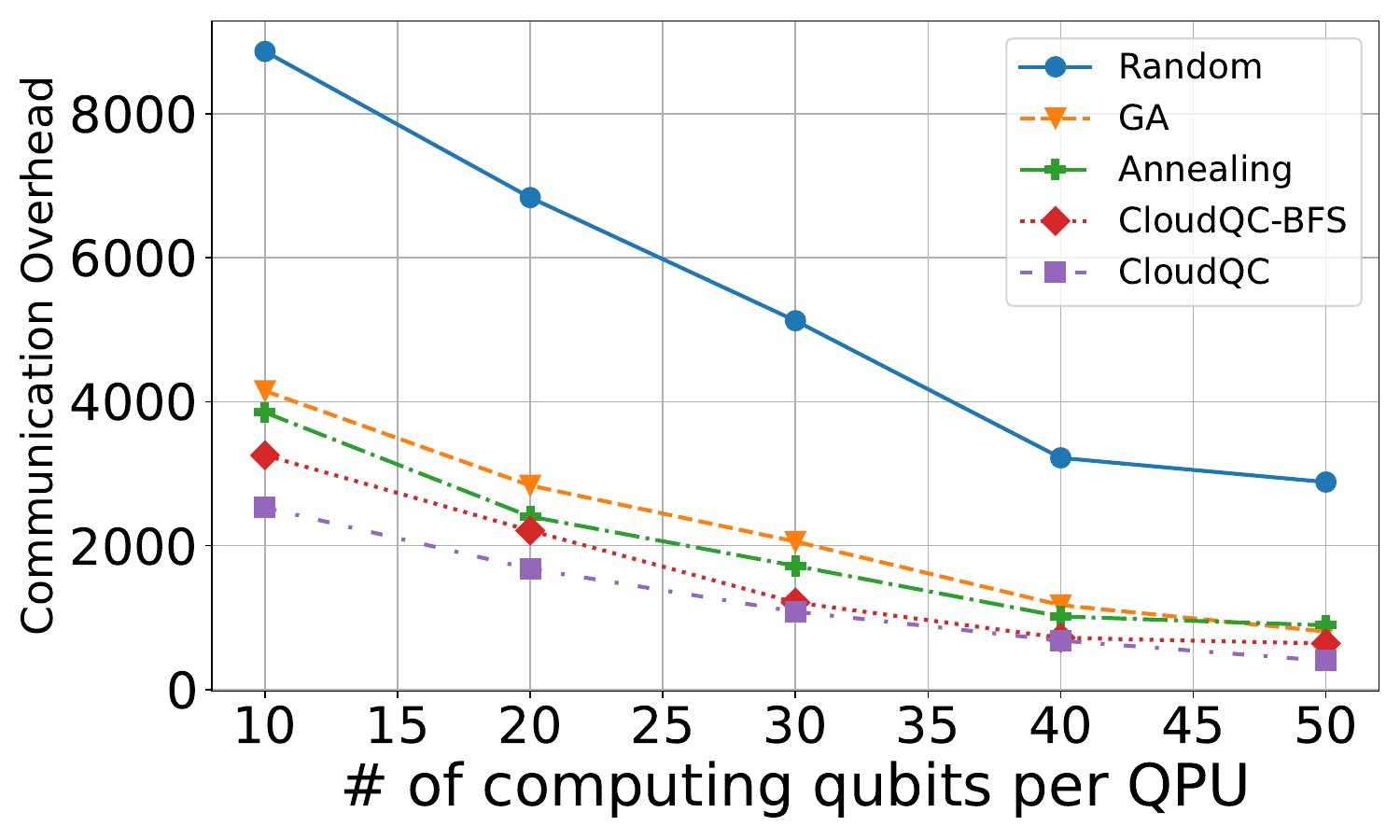}
    \vspace{-5ex}
    \caption{Overhead vs \# of computing qubits: multiplier\_n75}
    \label{fig-p}
  \end{minipage}\hfill
  \begin{minipage}[t]{0.24\textwidth}
    \includegraphics[width=\textwidth]{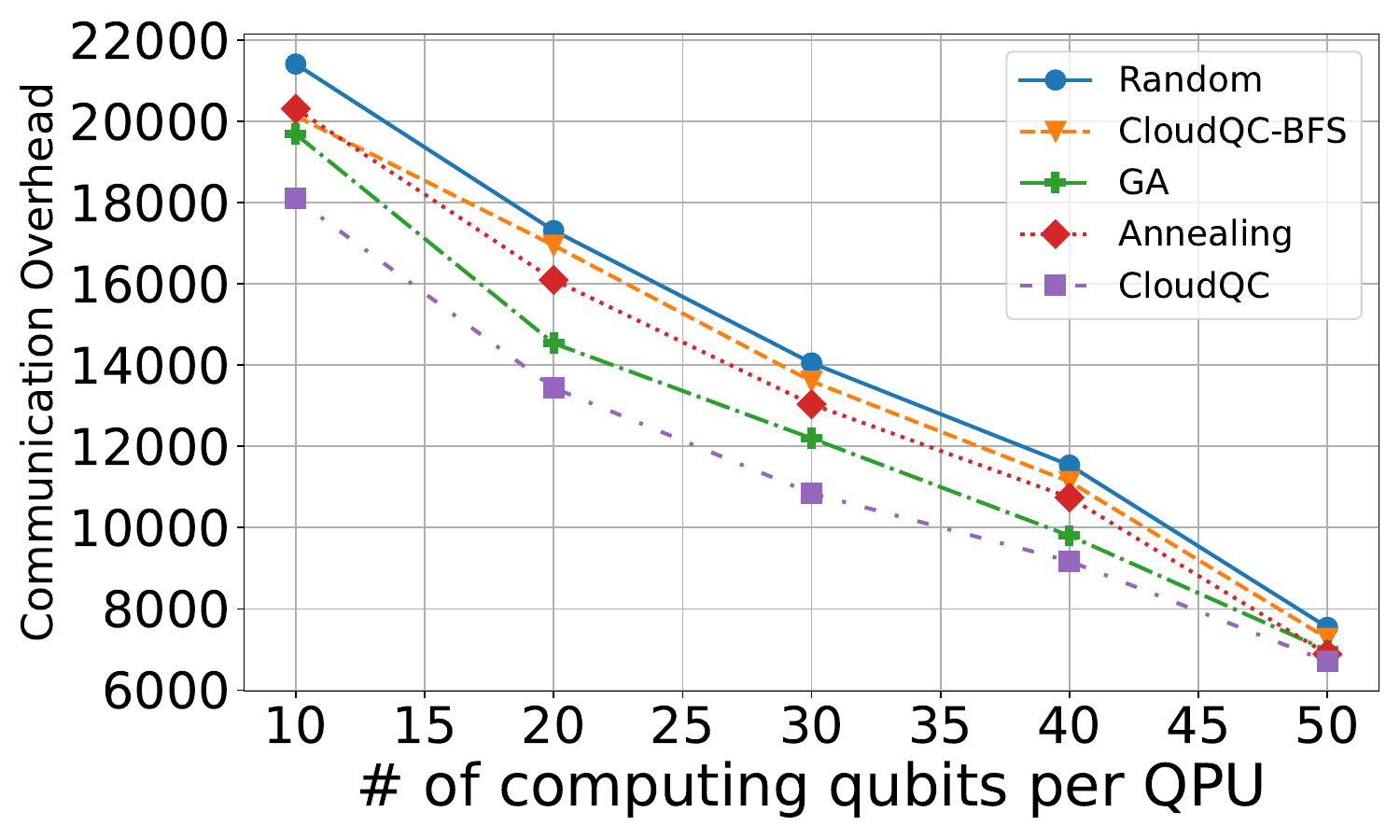}
    \vspace{-5ex}
    \caption{Overhead vs \# of computing qubits: QV\_n100}
    \label{fig-q}
  \end{minipage}
  \vspace{-3ex}
\end{figure*}

\begin{figure*}[!t]
  \centering
  \begin{minipage}[t]{0.24\textwidth}
    \includegraphics[width=\textwidth]{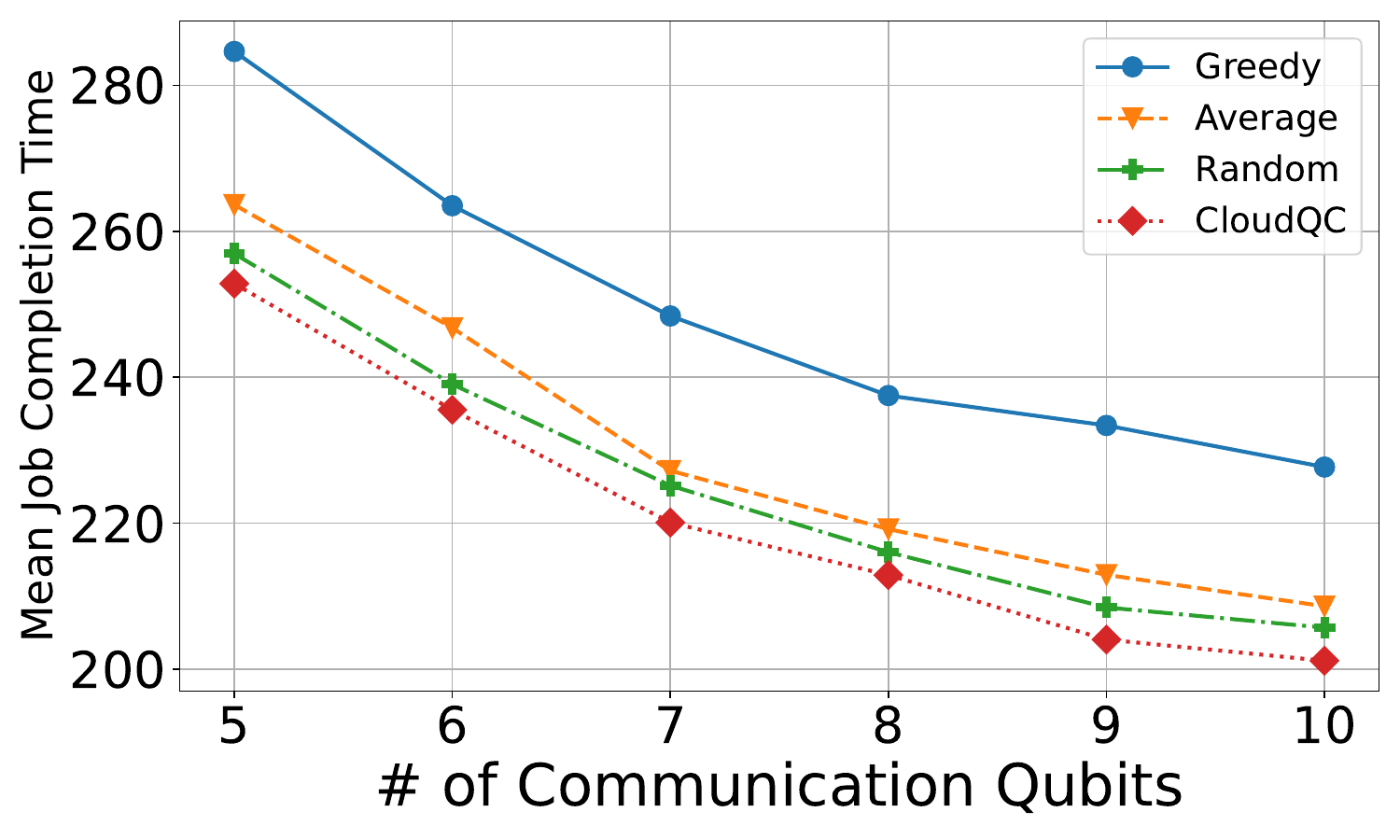}
    \vspace{-5ex}
    \caption{Job Completion Time vs \# of communication qubits: qugan\_n111}
    \label{fig-n1}
  \end{minipage}\hfill
  \begin{minipage}[t]{0.24\textwidth}
    \includegraphics[width=\textwidth]{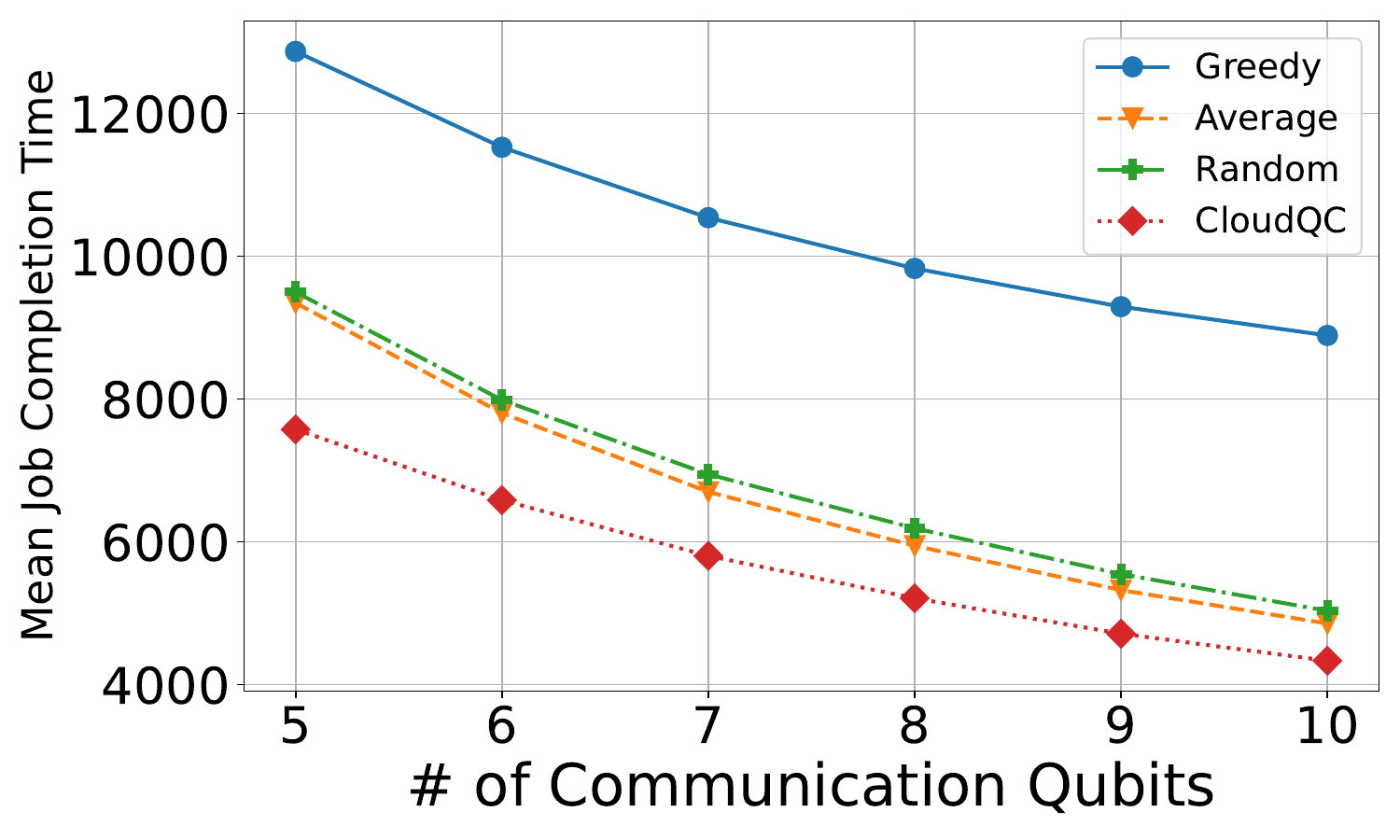}
    \vspace{-5ex}
    \caption{Job Completion Time vs \# of communication qubits: qft\_n160}
    \label{fig-d1}
  \end{minipage}\hfill
  \begin{minipage}[t]{0.24\textwidth}
    \includegraphics[width=\textwidth]{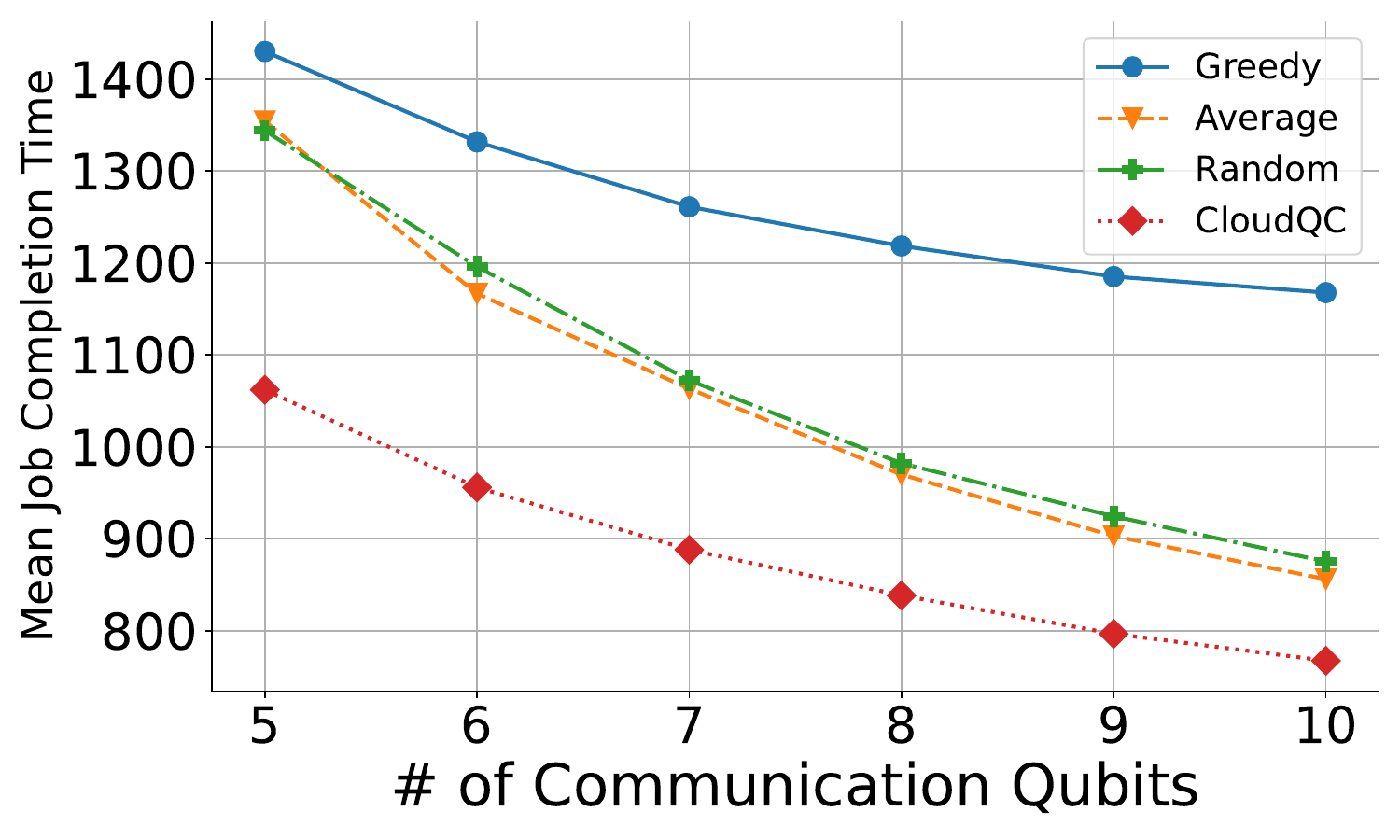}
    \vspace{-5ex}
    \caption{Job Completion Time vs \# of communication qubits: multiplier\_n75}
    \label{fig-p1}
  \end{minipage}\hfill
  \begin{minipage}[t]{0.24\textwidth}
    \includegraphics[width=\textwidth]{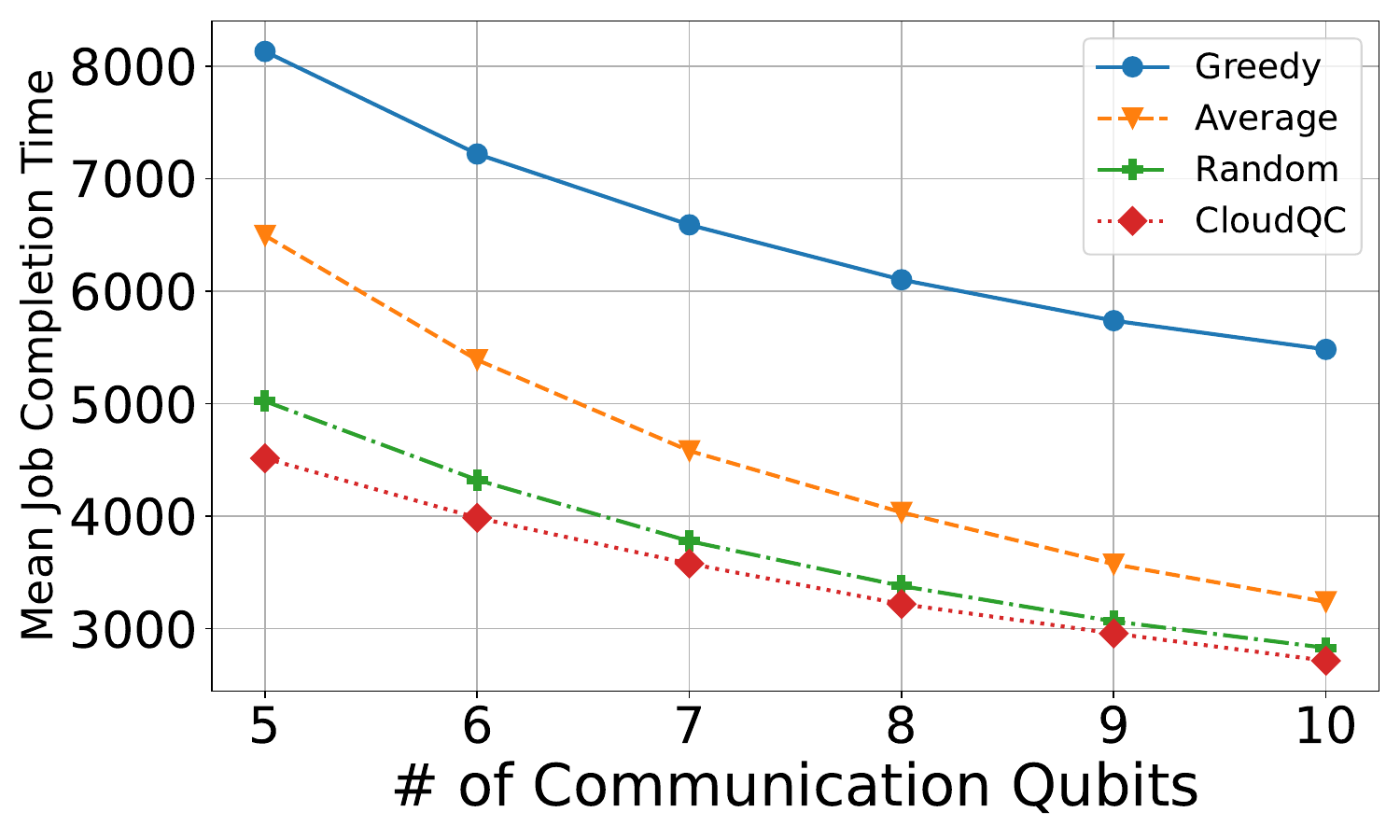}
    \vspace{-5ex}
    \caption{Job Completion Time vs \# of communication qubits: QV\_n100}
    \label{fig-q1}
  \end{minipage}
  \vspace{-3ex}
\end{figure*}

\begin{figure*}[!t]
  %\vspace{-3ex}
  \centering
  \begin{minipage}[t]{0.24\textwidth}
    \includegraphics[width=\textwidth]{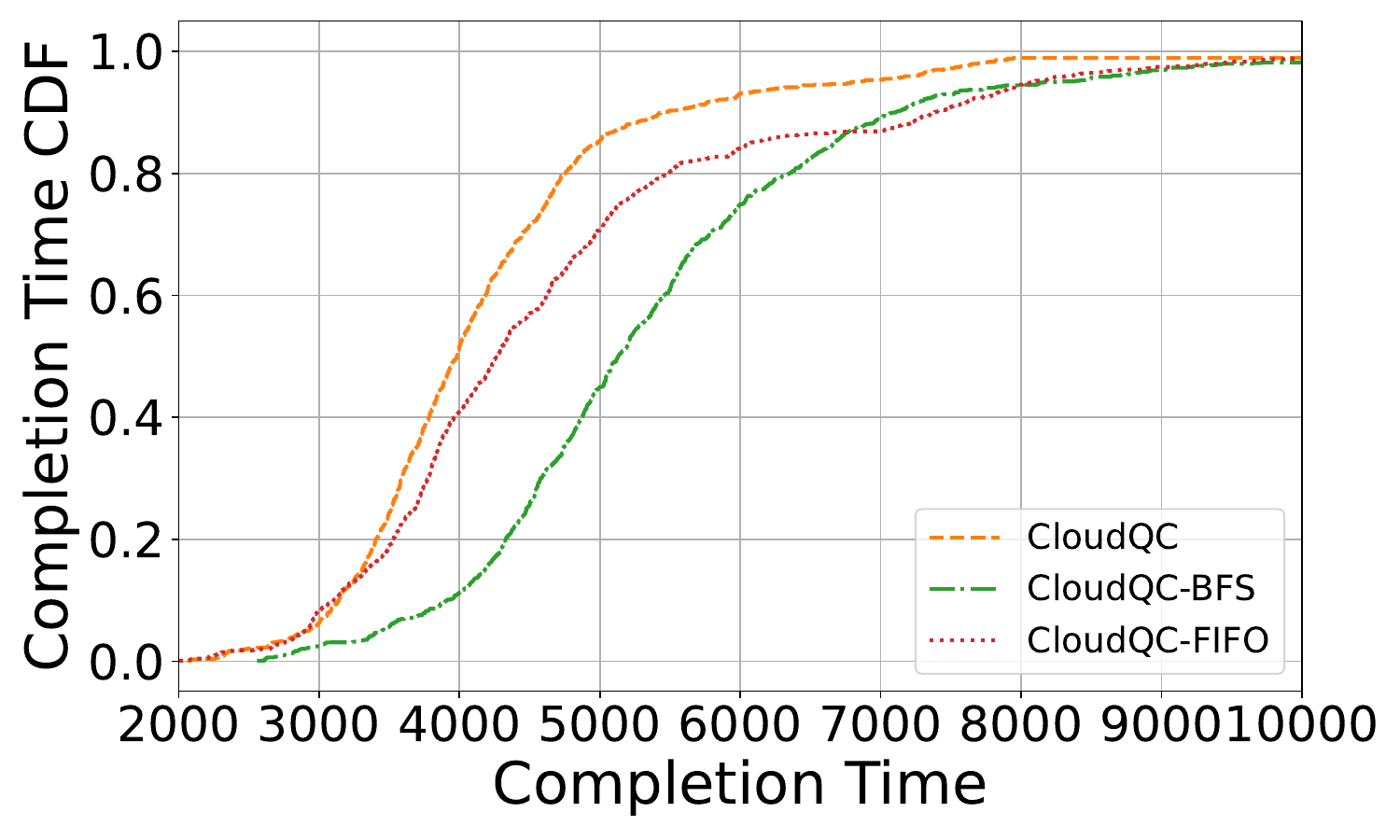}
    \vspace{-5ex}
    \caption{Job Completion CDF Time with Mixed Workloads}
    \label{fig-n3}
  \end{minipage}\hfill
  \begin{minipage}[t]{0.24\textwidth}
    \includegraphics[width=\textwidth]{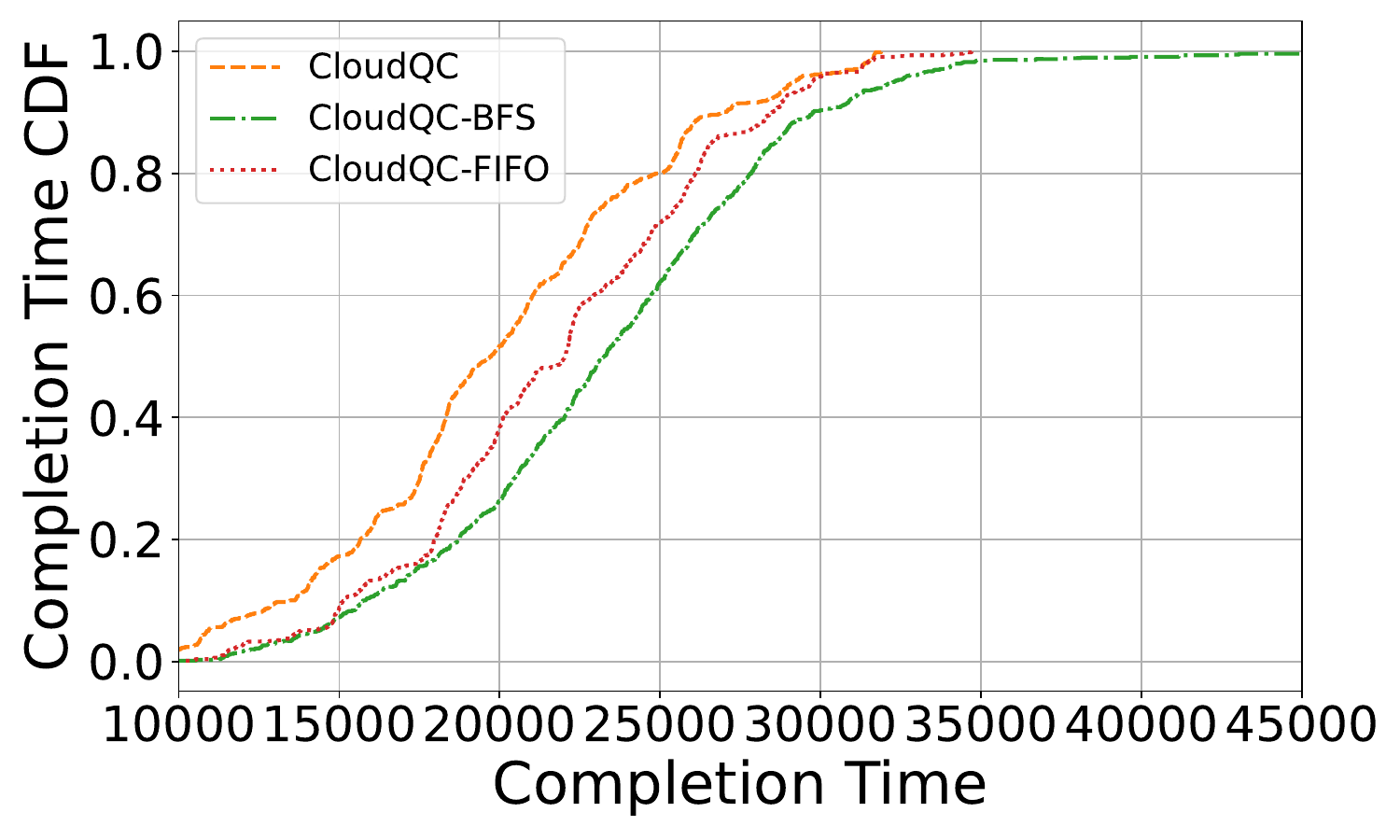}
    \vspace{-5ex}
    \caption{Job Completion Time CDF with QFT Workloads}
    \label{fig-d3}
  \end{minipage}\hfill
  \begin{minipage}[t]{0.24\textwidth}
    \includegraphics[width=\textwidth]{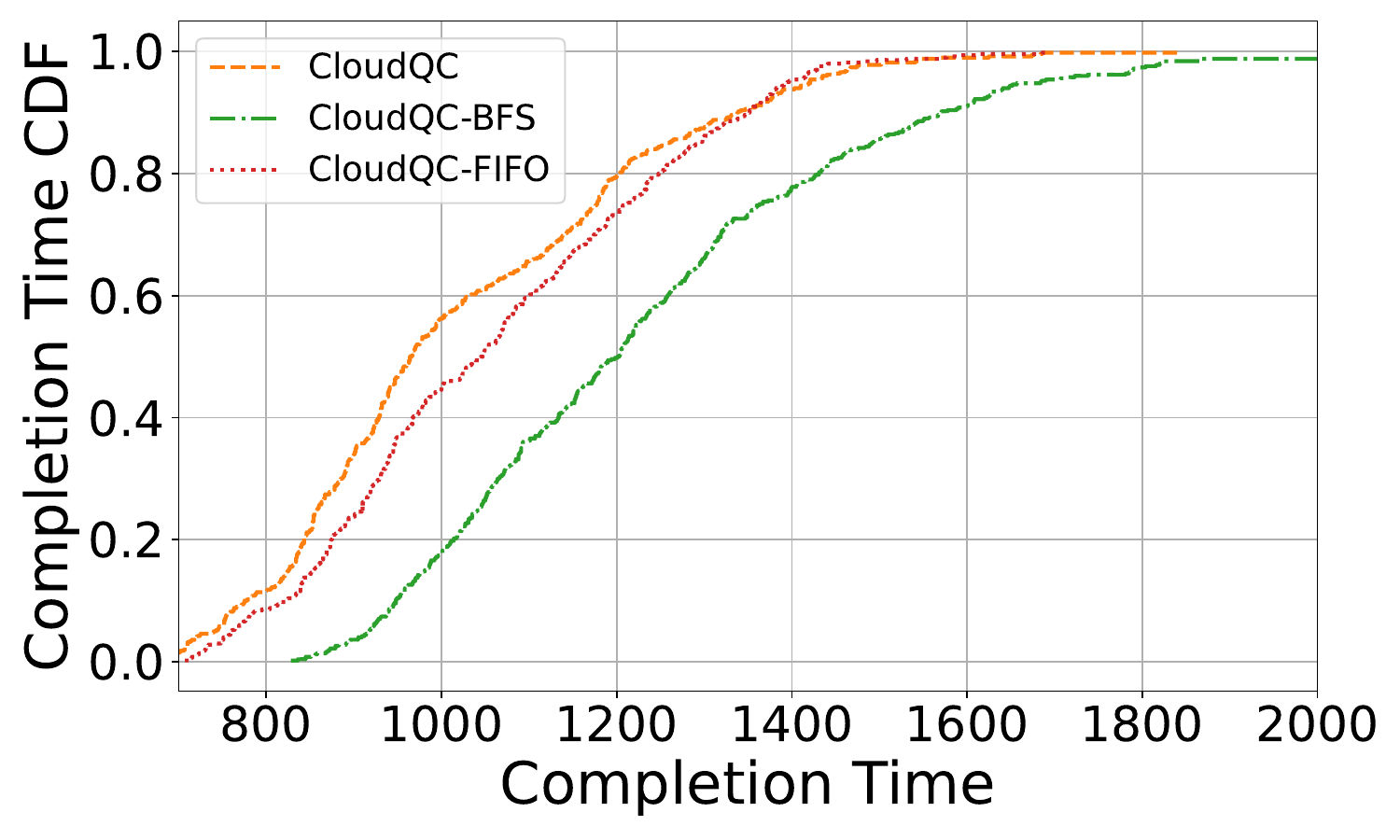}
    \vspace{-5ex}
    \caption{Job Completion Time CDF with Qugan Workloads}
    \label{fig-p3}
  \end{minipage}\hfill
  \begin{minipage}[t]{0.24\textwidth}
    \includegraphics[width=\textwidth]{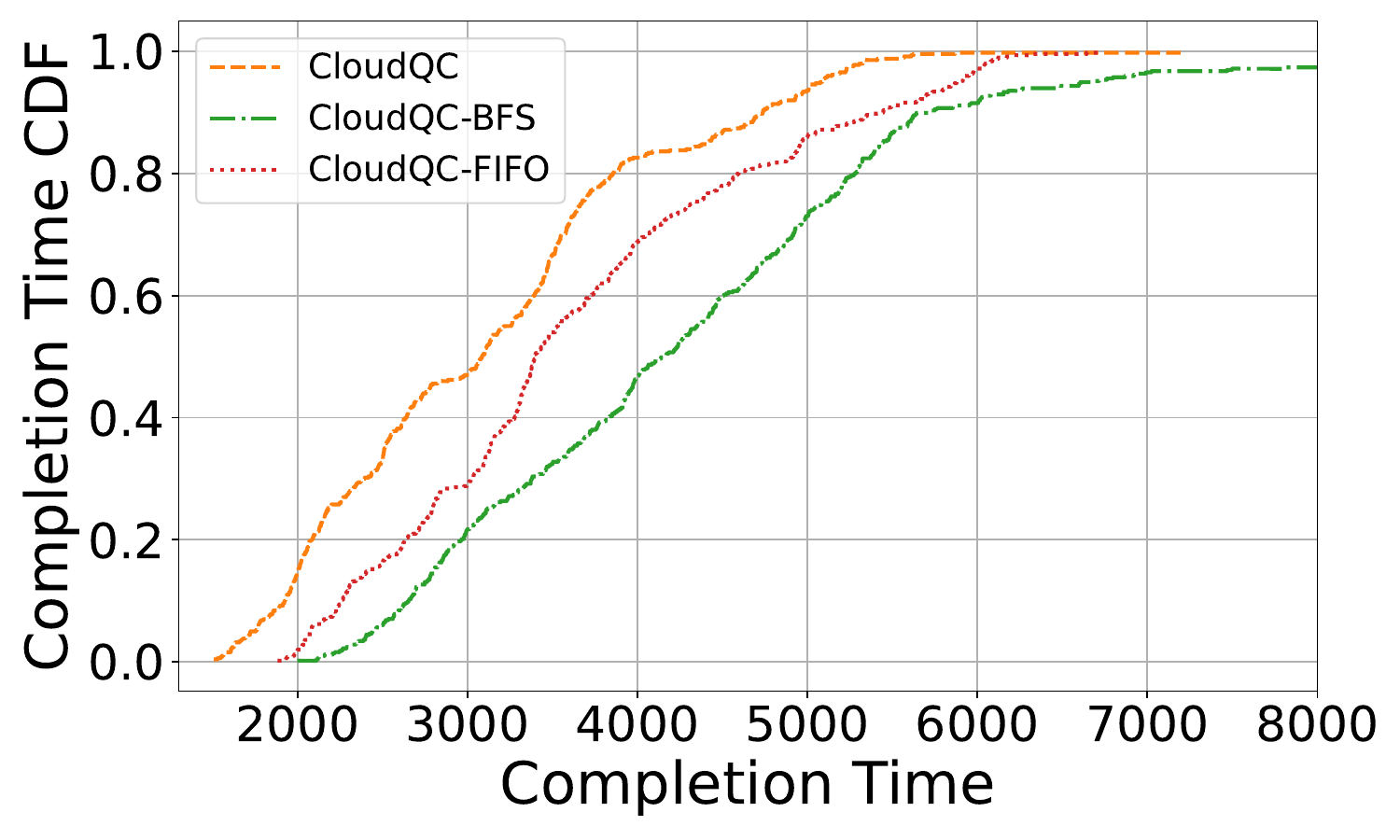}
    \vspace{-5ex}
    \caption{Job Completion Time CDF with Arithmetic Workloads}
    \label{fig-q3}
  \end{minipage}
  \vspace{-3ex}
\end{figure*}

\vspace{-1ex}
\subsection{Evaluation on Circuit Placement}
\vspace{-1ex}
% As shown previously, our circuit placement problem is a QAP problem by nature. We select two representatives of heuristics that have been widely used.
% \subsubsection{Baselines}
We first evaluate how CloudQC performs on placing single circuits. The metric is the communication cost $\sum_{i=1}^n \sum_{j=1}^n D_{ij}^{k} C_{\pi(q_i^{k}) \pi(q_j^{k})}$(In single circuits case we omit $x_k$ term) defined in Section III. We use the following baselines in comparison for circuit placement. 
\begin{itemize}
    \item Random Placement: It starts with a random node and does a random search to select a set of QPUs that meet computing constraints. %Then, it randomly assigns a qubit to the selected QPU.
    \item Simulated Annealing (SA): SA is a meta-heuristic that is widely used in optimization problems and placement in the cloud. Here, we use the strategy of a recent work~\cite{mao2023qubit}, which uses SA for qubit allocation for a single DQC job.
    \item Genetic Algorithm (GA)~\cite{holland1992genetic}: GA is also a meta-heuristic that is widely used in optimization problems.
    \item CloudQC-BFS: Also a method proposed by us. It differs from CloudQC in using a BFS search to find feasible QPU for each partition instead of community detection. 
    %\item CloudQC: Our proposed method first uses graph partition and then maps the partition to a set of QPUs found by the community detection algorithm. Eventually, we will map partitions to selected QPUs using our proposed mapping heuristic. 
\end{itemize}
\subsubsection{Circuit Placement for Single Circuit with Default Setting}
We show the results on circuit placement in Table~\ref{tab:experiment_results}. We can see that: 1) CloudQC (and CloudQC-BFS) significantly outperforms other baselines for most circuits. For cc\_n64, bv\_n70, knn\_n96, cat\_n65, cat\_n130, ising\_n34 and ghz\_n127, CloudQC provides similar results to CloudQC-BFS. The reason is that these circuits have fewer 2-qubit gates, and graph partition contributes most to the final result. %Thus, the two methods give almost the same result. 
%The remote interaction graph of these methods is simple, and the BFS search and our proposed methods gave almost the same behavior. 
However in larger circuits with more complicated inter-QPU interactions, such as qft\_n63, qft\_n160, multiplier\_n75, and swap\_test\_n115, CloudQC performs significantly better than CloudQC-BFS, which shows the effectiveness of the proposed mapping method and community detection. 
%Our community detection method works well in that it can utilize the topology of the quantum cloud and select a list of tightly connected QPUs, which gives the optimization space for later mapping. Our simple mapping heuristic tends to place two communication-intensive partitions closer, thus efficiently reducing communication overhead. 
%2) Among all benchmarks, random placement gives the worst placement result since it ignores the circuits' communication pattern. 
3) SA works better than random placement but much worse than CloudQC. The reason is that the performance of SA highly depends on the initial placement. We also observed a long running time ($>$1 hour) of SA and GA while CloudQC finishes within 2 minutes for most circuits.%We can see that simply adopting SA fails to provide an ideal solution without an optimized initial solution.

\subsubsection{Effect of Number of Computing Qubits}
We vary the number of computing qubits on each QPU by 20, 30, 40, and 50. Due to the page limit, we select the following representative quantum circuits as benchmarks: \textit{qv\_n100}, \textit{multiplier\_n75}, \textit{qft\_n160}, and \textit{qugan\_n111}. %These circuits are chosen because they are the most challenging to place, having a higher number of qubits and more 2-qubit gates. We summarised the effect of the number of computing qubits on circuit placement 
We show the results in Figs.~\ref{fig-n} to \ref{fig-q}. We can see CloudQC performs the best for these circuits, and CloudQC-BFS also performs better than other baselines. 
%but worse than ours since it uses a similar partition strategy to ours but uses a search-based method to select QPUs. 
Although it also tends to select close connected QPUs it may fail to capture the topology information of the cloud. SA and GA perform worse than our methods in that it fails to capture the communication information between different qubits in the circuit. GA performs better than SA, but we also observe a much longer running time. 

\begin{figure*}[!t]
  \centering
  \begin{minipage}[t]{0.24\textwidth}
    \includegraphics[width=\textwidth]{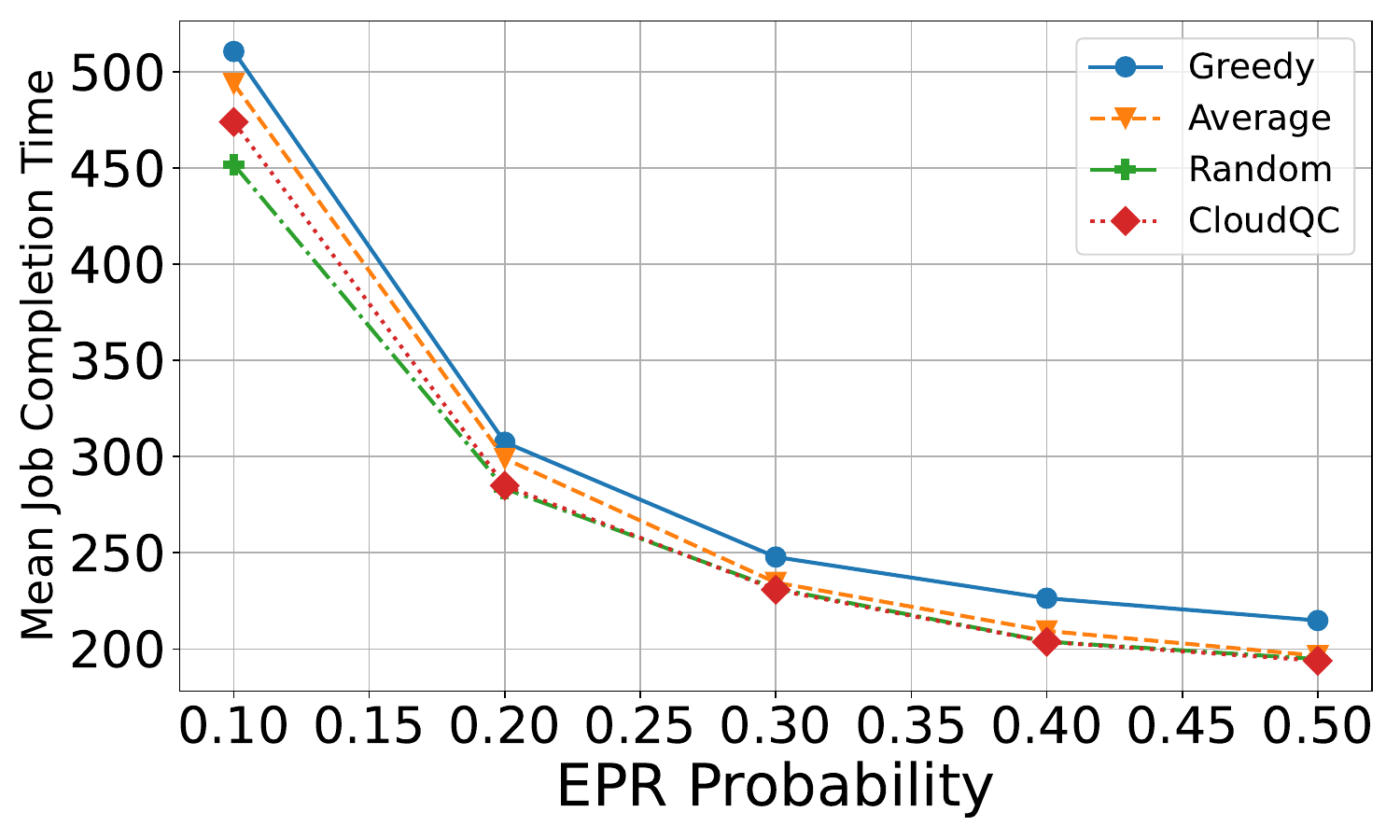}
    \vspace{-5ex}
    \caption{Job Completion Time vs EPR probability: qugan\_n111}
    \label{fig-n2}
  \end{minipage}\hfill
  \begin{minipage}[t]{0.24\textwidth}
    \includegraphics[width=\textwidth]{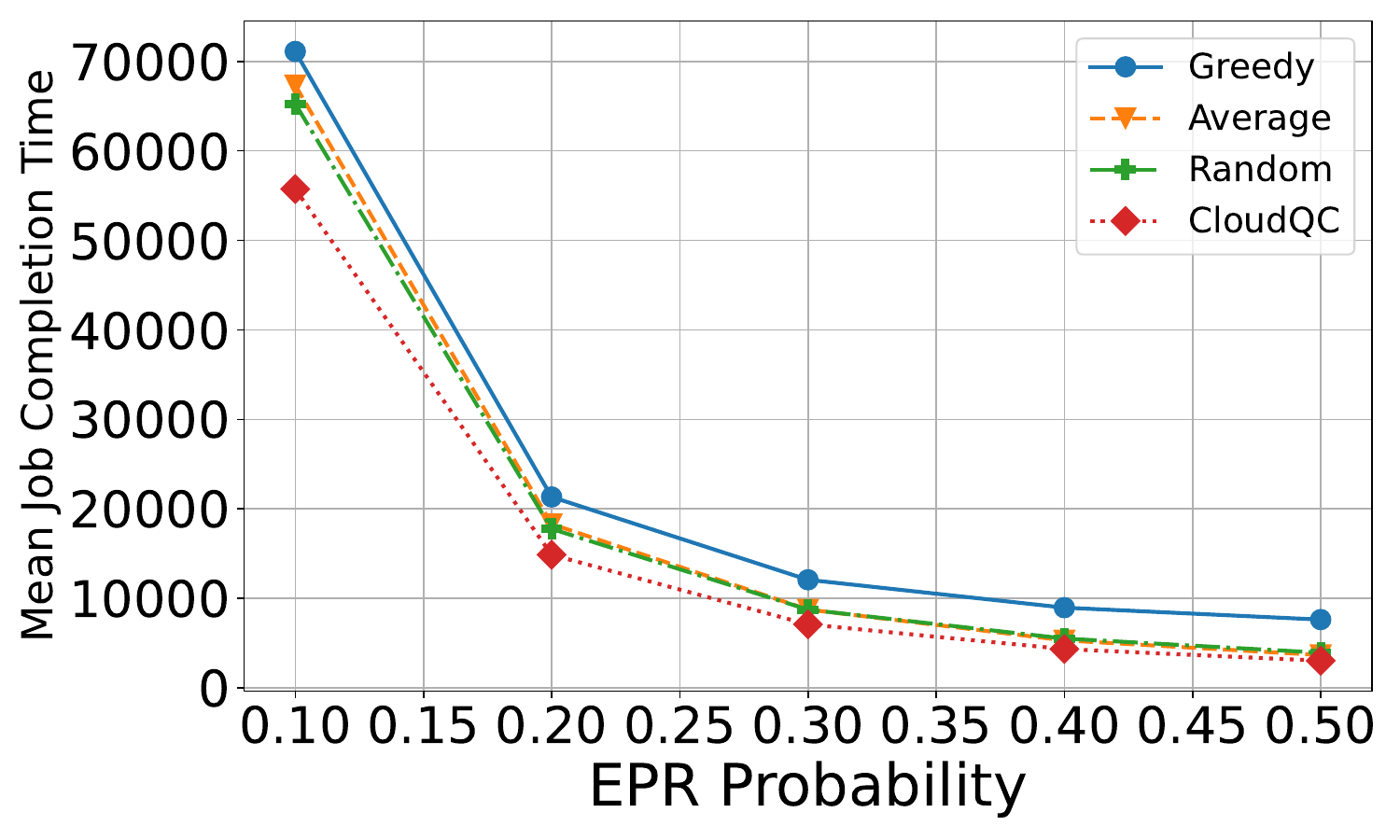}
    \vspace{-5ex}
    \caption{Job Completion Time vs EPR probability: qft\_n160}
    \label{fig-d2}
  \end{minipage}\hfill
  \begin{minipage}[t]{0.24\textwidth}
    \includegraphics[width=\textwidth]{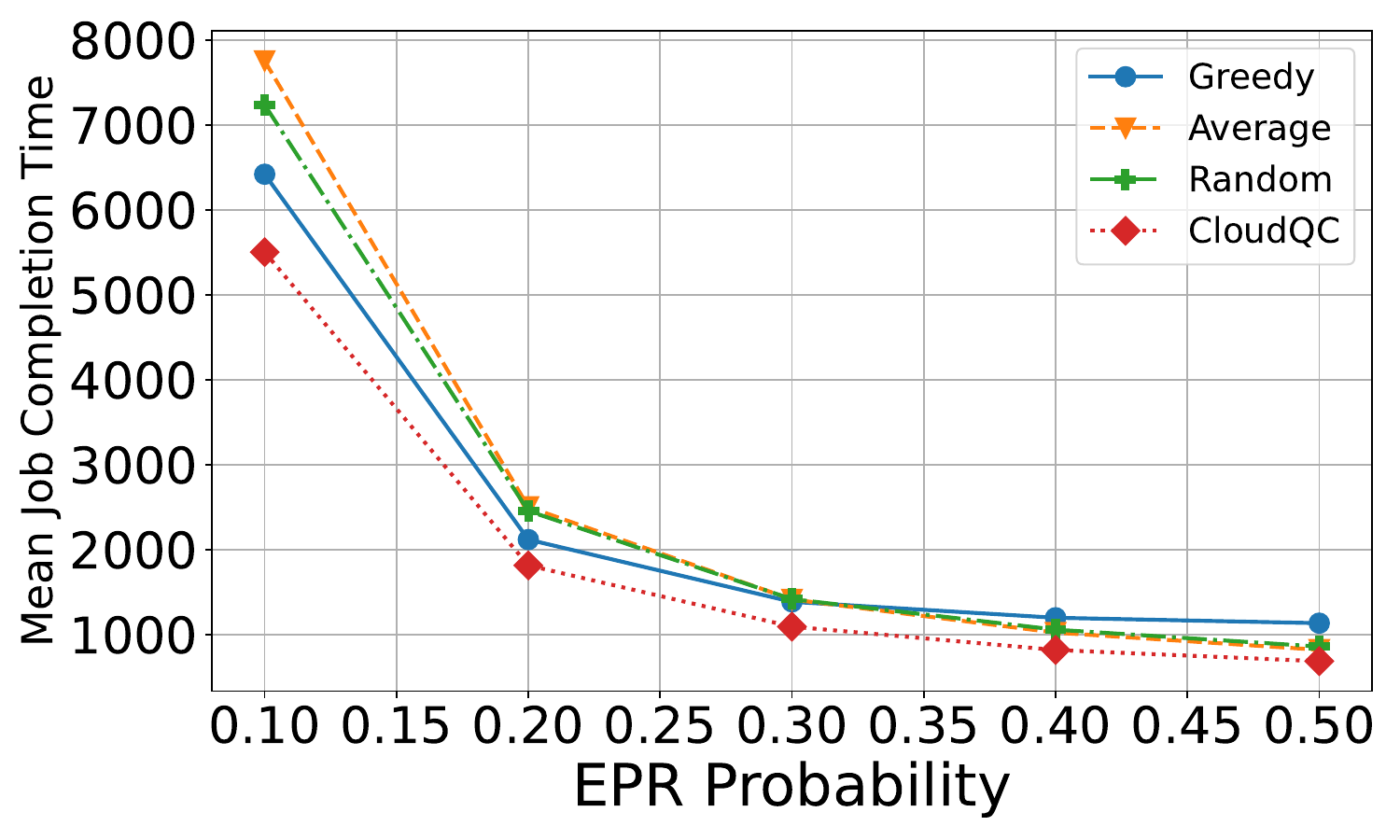}
    \vspace{-5ex}
    \caption{Job Completion Time vs EPR probability:multiplier\_n75}
    \label{fig-p2}
  \end{minipage}\hfill
  \begin{minipage}[t]{0.24\textwidth}
    \includegraphics[width=\textwidth]{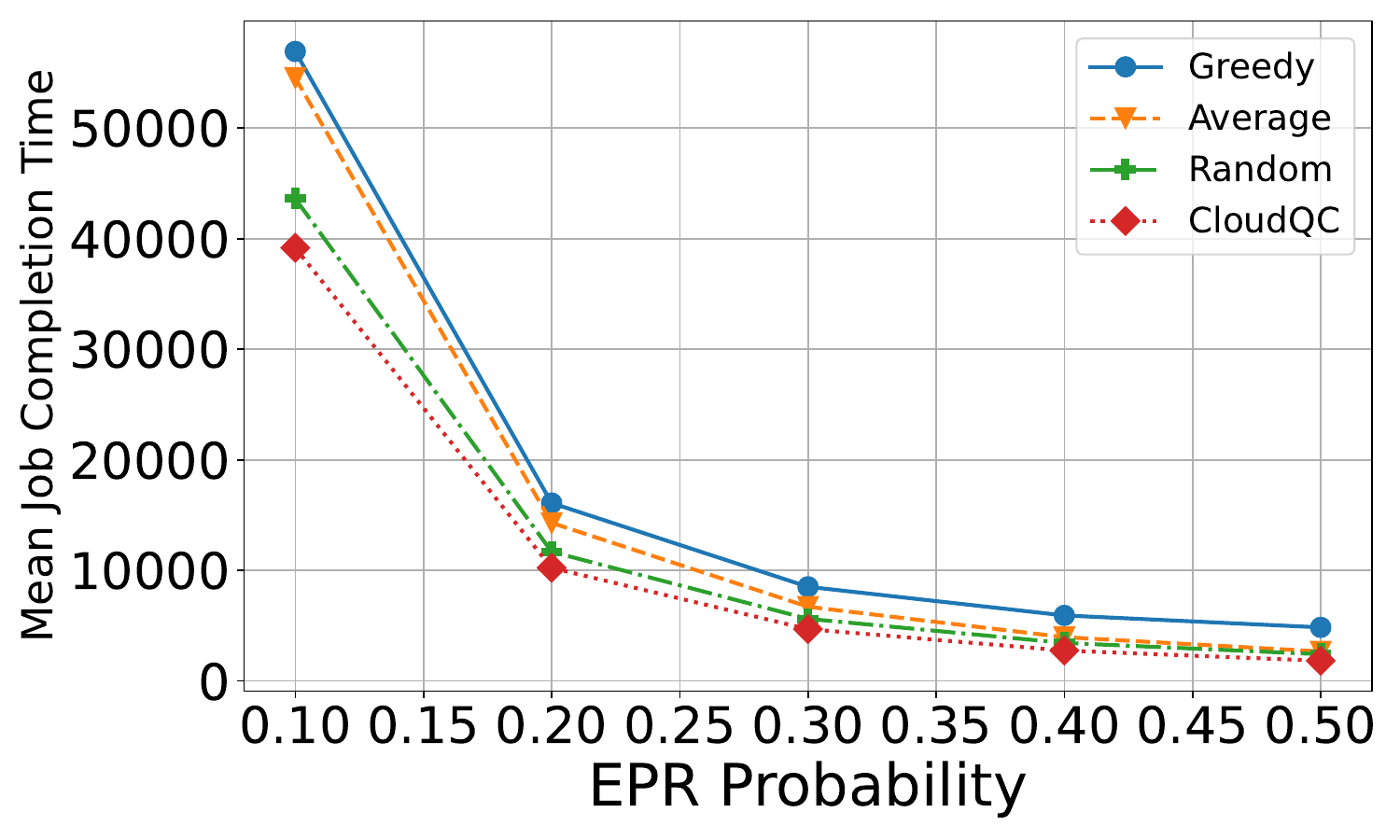}
    \vspace{-5ex}
    \caption{Job Completion Time vs EPR probability: QV\_n100}
    \label{fig-q2}
  \end{minipage}
  \vspace{-3ex}
\end{figure*}
\vspace{-1.5ex}

\subsection{Evaluation on Network Scheduling}
% multiplier, qft, knn, qugan
We compare the following methods to evaluate the flowing scheduling in CloudQC. 
\begin{itemize}
    %\item Priority: Our proposed network scheduling strategy, which is proposed in section V. C
    \item Greedy: It always allocates the maximum resources to the remote operation with the highest priority.
    \item Average: It distributes communication resources evenly among all remote operations.
    \item Random: %Each unit of communication resources is allocated randomly, giving 
    Each remote operation has an equal probability of receiving communication resources.
\end{itemize}
\subsubsection{Network Scheduling with Default Setting}
\begin{figure}[!t]
\setlength{\abovecaptionskip}{0.cm}
\centerline{\includegraphics[width=0.45\textwidth]{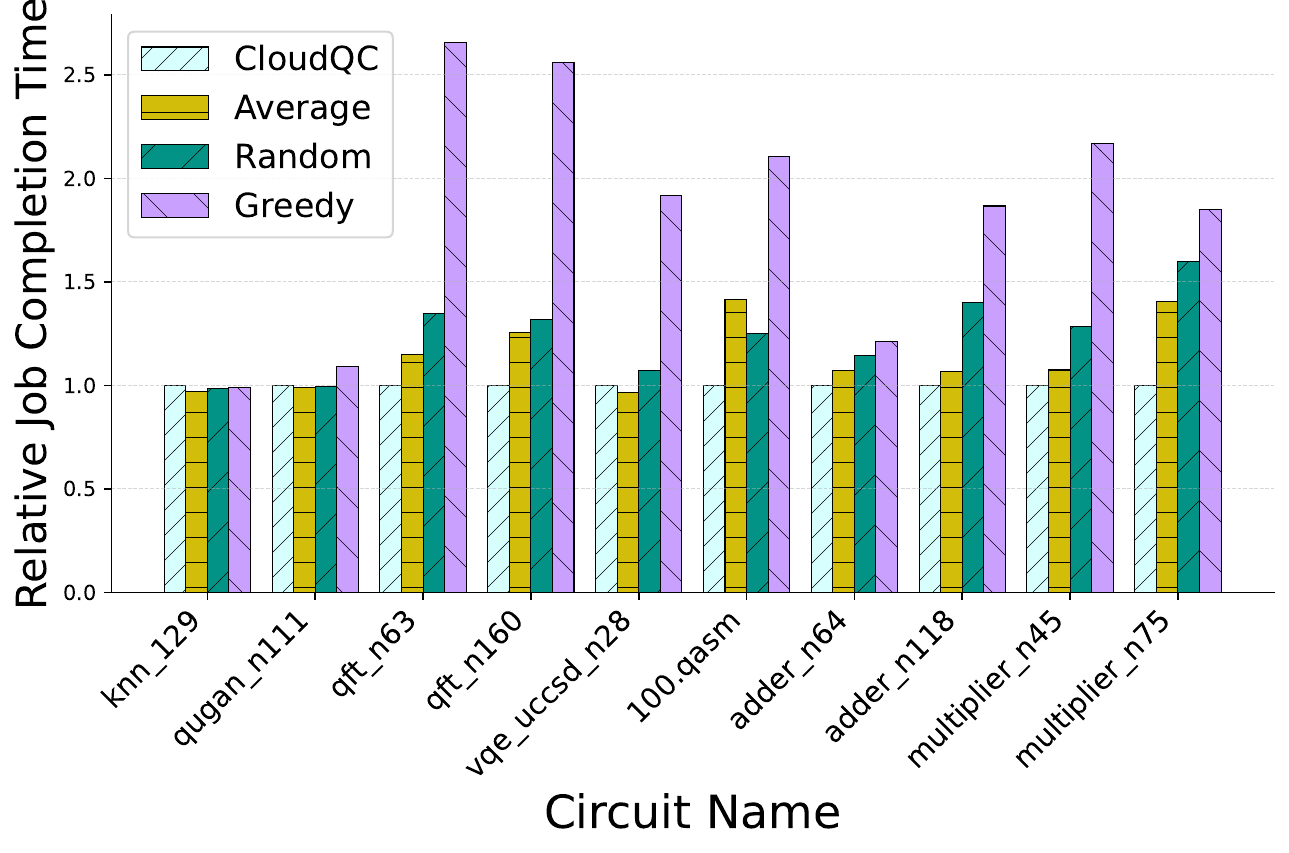}}
\vspace{-2ex}
\caption{Network Scheduling with Different Methods with Default Setting}
\vspace{-3ex}
\label{fig:scheduling}
\end{figure}

Figure \ref{fig:scheduling} illustrates the network scheduling results with default settings. 
%Using the same set of circuits as in previous sections, the results show that our methods 
CloudQC significantly achieves the least job completion time for most quantum circuits especially with more complex and random structures such as Quantum Fourier Transform (QFT), Multiplier, and Quantum Volume (QV) circuits. This is due to their ability to utilize the DAG topology. For circuits without complex DAG topology, such as BV, Qugan, KNN, and Swap test, CloudQC performs similarly to others. Greedy has the worst job completion time.
%method performs the worst, particularly in QFT, Multiplier, and QV circuits, due to its inefficient resource allocation.

\subsubsection{Effect of Number of Communication Qubits}
We vary the number of communication qubits from 5 to 10 and show the corresponding results in Figs.~\ref{fig-n1} to \ref{fig-q1}. We use qugan\_n111, qft\_n63, multiplier\_n75 and QV\_n100 as representative benchmarks. From all results, we can see the number of communication qubits has a large impact on the completion time. Similar to previous observations for circuits with a more complicated structure, CloudQC achieves a significantly shorter completion time compared to other methods.
%And in circuits such as qugan\_n111, different methods do not show a big difference. 
\subsubsection{Effect of EPR success Probability}
We vary the EPR success probability from 0.1 to 0.5, and the corresponding results are shown from Figs.\ref{fig-n2} to \ref{fig-q2}. 
%Using the same set of circuits as in previous evaluations, 
We observe that increasing the EPR probability decreases job completion time. %Notably, there is a significant reduction in completion time when EPR probability is increased from 0.1 to 0.2. 
Across all scenarios, CloudQC consistently achieves shorter job completion time except for one data point (probability 0.1 for qugan\_n111). %other methods, while the Random method performs similarly to the Average method, and the Greedy allocation performs the worst. 
With these results, we can see that improving the success rate of EPR pairs will be a crucial step in the future DQC hardware. 

\vspace{-1ex}
\subsection{Evaluation on Multi-Tennant Settings}
\vspace{-1ex}
% \begin{figure}[!t]
% \setlength{\abovecaptionskip}{0.cm}
% \centerline{\includegraphics[width=0.49\textwidth]{cdf_batch_completion_time_all_experiments.pdf}}

% \caption{Flow Scheduling with Different Methods with Default Setting}

% \label{circuit_exmple}

% \end

\subsubsection{Evaluation Setting}
For multi-tenant experiments, we use the following workloads of different types of quantum circuits with different numbers of qubits:%. We use the following workloads:
\begin{itemize}
    \item Mixed workloads: It contains different types of quantum circuits with different numbers of qubits: knn\_n129, qugan\_n111, qugan\_n71, qft\_n63, multiplier\_n45, multiplier\_n75.
    \item QFT workloads: It contains QFT circuits with different numbers of qubits: qft\_n29, qft\_n63, qft\_n100.
    \item Qugan workloads: It contains Qugan circuits with different numbers of qubits:qugan\_n39, qugan\_n71, qugan\_n111.
    \item Arithmetic workloads: It contains different types of quantum arithmetic circuits with different numbers of qubits: adder\_n64, adder\_n118, multiplier\_n45, multiplier\_n75
\end{itemize}
%To evaluate our framework's capability of scheduling in time and space, 
We compare CloudQC with the following methods:
\begin{itemize}
    %\item Default: Our proposed CloudQC framework, which uses the proposed batch manager to determine the process order and GCM method for circuit placement, and our proposed priority-based network scheduling method. 
    \item CloudQC-BFS: It uses the same batch manager as CloudQC but uses the BFS method in each circuit placement.
    % since it performs only less than our proposed method and shows a much shorter running time than the annealing and genetic algorithm method. 
    \item CloudQC-FIFO: It uses the same placement and network scheduling methods  as CloudQC but a First-In-First-Out processing order for batched jobs.%, which means we will always try to find placement for the first circuit that has not been executed in the batch.  
\end{itemize}
%We use the same setting as that in the previous single-circuit experiments. 
For each workload, 
%we form a batch with 20 circuits selected randomly and run the experiments 20 times. 
we generate 50 batches, each of which includes 20 circuits selected randomly from the workload. Each batch is run 20 times with different network topologies. 
%in total, and for each workload, we run 1000 experiments. 

We show the results from Figs.~\ref{fig-n3} to \ref{fig-q3} as the CDF of the job completion time. We find that CloudQC performs better than other methods, followed by CloudQC-FIFO. In mixed workloads, CloudQC finishes around 88\% in around 5000 units of time, and CloudQC-FIFO only finishes around 70\% at the same time. In circuits with complicated patterns and longer depths, we can see CloudQC still performs significantly better than CloudQC-BFS and CloudQC-FIFO. However for circuits with shorter depths(Qugan), the differences are small. %BFS performs worst in that it does not find placement for each circuit in space in that it leaves many 'fragments' in the quantum cloud, 
CloudQC-BFS does not perform well in multi-tenant scenarios although in placing single circuits, it achieves the smaller cost. Hence we find that CloudQC is the best choice for both single-circuit and multi-circuit placement and network scheduling. 
%a placement algorithm that works well for a single circuit does not necessarily work well in multi-tenant settings., which is also consistent with our previous methods. FIFO performs worse in our method in that it does not schedule circuits well in time, and methods to consider various factors are needed. 

\vspace{-1ex}
\section{Related Work}
\vspace{-1ex}
\textbf{Distributed Quantum Computing}. Distributed quantum computing has been studied in recent years. Some of the works focus on qubit allocation and compilers~\cite{mao2023qubit,baker2020time,andres2019automated,diadamo2021distributed,wu2022autocomm}. Mao \textit{et al}. \cite{mao2023qubit} proposed to use a hybrid simulated annealing algorithm to determine the qubits allocation in distributed quantum computing. 
%Instead of using a weighted graph, 
Andres-Martinez and Heunen \cite{andres2019automated}  used partitioning hypergraph to minimize communication costs. Baker \textit{et al}. \cite{baker_time-sliced_2020} used remote SWAP gates to replace all remote CX gates in distributed quantum programs and obtain higher throughput. %\cite{diadamo_distributed_2021} proposed to utilize multi-qubit control block for VQE algorithm. 
Autucomm \cite{wu2022autocomm} identified burst communication patterns in DQC  and determined the best choice for using the cat-entangler method or teleportation method. These works mainly focus on optimizing communication of a single circuit and do not take the probabilistic nature of quantum communication into consideration. %and thus are orthogonal into consideration. 

\textbf{Quantum Multi-programming and Quantum Cloud}.The idea of quantum multi-programming was proposed in \cite{das2019case}; it considers fairness when allocating multiple quantum programs on one single QPU. 
\cite{niu2023enabling} use similar ideas and evaluate their methods on various benchmarks on real devices. Liu \textit{et al}. \cite{liu2023data} proposed a quantum data center architecture composed of quantum random access memory and quantum networks and visioned three quantum applications. Ravi \textit{et al}.\cite{ravi2021quantum} also surveyed various applications and resource utilization problems in a cloud environment for quantum computing.

 %\cite{liu2021qucloud} used a community detection algorithm to find highly connected components to allocate quantum programs and consider inter-program SWAP gate to further improve throughput and reduce overhead. 
\vspace{-1ex}
\section{Conclusion}
\vspace{-1ex}
This work presents a network-aware framework for DQC in a multi-tenant quantum cloud. The framework, called CloudQC, consists of two critical components: circuit placement and network scheduling. 
CloudQC is the first work to consider the placement and network scheduling for multiple concurrent DQC circuits. 
In addition, it incorporates the probabilistic nature of quantum networks to allow redundant network resources for important quantum gates in each circuit avoiding backlogs of later circuits and reducing the job completion time. 
%For circuit placement, we introduce a heuristic that first distributes a quantum circuit and then assigns it to QPU candidates, selected using a modularity-based community detection algorithm, to map partitions to QPUs effectively. In the network scheduling component, we address the communication resource allocation problem in distributed quantum computing. Our goal is to minimize the execution time of quantum circuits, a key performance metric in this domain. We propose a straightforward heuristic for resource allocation that captures the structure of the distributed quantum circuit. To validate our approach, 
%We conduct experiments to simulate a practical quantum cloud environment with real-world large-scale quantum circuits as benchmarks for comprehensive evaluation. 
The simulation results show that CloudQC significantly reduces the job completion time and cost for both single-circuit and multiple-circuit placement and scheduling. 
We believe our work is a valuable step in quantum computing and networks as it envisions a practical quantum cloud infrastructure that will surely emerge in the near future. 
%, akin to today’s cloud and data centers. We explore this advanced infrastructure by addressing the necessary execution framework and highlighting the differences from traditional cloud computing. Our efforts aim to inspire further research and development towards realizing large-scale, practical, distributed quantum computing.
%\vspace{-1ex}

%\vspace{-2ex}

\bibliographystyle{IEEEtran}
\bibliography{ref}

\end{document}